\documentclass[12pt]{article}

\usepackage{scicite,adjustbox}
\usepackage{etoc}
\usepackage{times}
\usepackage{draftwatermark}
\SetWatermarkText{ }
\SetWatermarkScale{1.2}
\SetWatermarkColor[gray]{0.95}

\usepackage{graphicx} 
\usepackage{multirow} 
\usepackage{hyperref}
\usepackage{amsmath} 
\usepackage{amssymb} 
\newcommand{\x}{\mathbf{x}}
\usepackage{floatrow}
\usepackage{setspace}
\newfloatcommand{capbtabbox}{table}[][\FBwidth]

\newenvironment{sciabstract}{%
\begin{quote} \bf}
{\end{quote}}

\newcounter{lastnote}

\title{Transferable Generative Models Bridge Femtosecond to Nanosecond Time-Step Molecular Dynamics}

\author
{Juan Viguera Diez,$^{1,2}$, Mathias Schreiner,$^{1}$ and Simon Olsson$^{1,\ast}$\\
\\
\normalsize{$^{1}$Department of Computer Science and Engineering,}\\ \normalsize{Chalmers University of Technology and University of Gothenburg,} \\ \normalsize{SE-41296 Gothenburg, Sweden}\\
\normalsize{$^{2}$Molecular AI, Discovery Sciences, R\&D, }\\\normalsize{AstraZeneca Gothenburg,}  \\ \normalsize{Pepparedsleden 1, 431 50 Mölndal, Sweden.}\\
\\
\normalsize{$^\ast$To whom correspondence should be addressed; E-mail:  simonols@chalmers.se.}\\
}

\date{}

\begin{document} 


\baselineskip24pt


\maketitle

\vspace{-0.75cm}
\begin{sciabstract}
Understanding molecular structure, dynamics, and reactivity requires bridging processes that occur across widely separated timescales. Conventional molecular dynamics simulations provide atomistic resolution, but their femtosecond time steps limit access to the slow conformational changes and relaxation processes that govern chemical function. Here, we introduce a deep generative modeling framework that accelerates sampling of molecular dynamics by four orders of magnitude while retaining physical realism. Applied to small organic molecules and peptides, the approach enables quantitative characterization of equilibrium ensembles and dynamical relaxation processes that were previously only accessible by costly brute-force simulation. Importantly, the method generalizes across chemical composition and system size, extrapolating to peptides larger than those used for training, and captures chemically meaningful transitions on extended timescales. By expanding the accessible range of molecular motions without sacrificing atomistic detail, this approach opens new opportunities for probing conformational landscapes, thermodynamics, and kinetics in systems central to chemistry and biophysics.
\end{sciabstract}





\section*{Introduction}
Many of the most important observables in statistical mechanics---such as the stability of a folded protein, the conformational transitions underlying allosteric regulation, or the unbinding rate of a drug from its target---are central to understanding chemical and biological function. These processes span timescales from nanoseconds to seconds, and while they are directly accessible through experiments such as spectroscopy \cite{PalmerIII2001} and single-molecule techniques \cite{Schuler2013}, their atomistic origins are often hidden.

Molecular dynamics (MD) offers a powerful complement to such experiments. By simulating the trajectories of atoms and molecules at atomic resolution, MD connects the fundamental interatomic forces that govern molecular motion to the statistical behavior observed in bulk. In this way, simulations provide a mechanistic bridge between microscopic physics and macroscopic phenomena \cite{Leimkuhler2015}.

Yet MD comes with a fundamental limitation. To ensure numerical stability, simulations must take time steps small enough to resolve the fastest motions in the system, such as bond and angle vibrations. This requirement restricts MD to femtosecond update steps, even though many processes of chemical and biological interest --- protein folding, conformational transitions, ligand binding --- unfold over microseconds to seconds. These processes are typically governed by rare transitions between metastable states \cite{Karplus2002}, creating a persistent gap between simulation and experiment that limits our ability to characterize slow molecular processes with statistical confidence \cite{schutte2009conformation}.

This challenge, known as the {\it `sampling problem’}, continues to inspire a growing array of strategies aiming at accelerating the observation of rare events. Most approaches either bias the underlying dynamics or simulate multiple coupled replicas in parallel, both designed to make infrequent transitions occur more often \cite{Henin2022}. Biasing methods rely on the definition of collective variables (CVs), that capture the progress of a process of interest \cite{Laio2002,Torrie1977,Grubmuller1995}. However, identifying suitable CVs for complex, high-dimensional systems remains difficult: variables that accelerate one process may obscure others, and their design has become a discipline in its own right with numerous options \cite{Branduardi2007} including ones derived using machine learning-based strategies \cite{PrezHernndez2013,Tiwary2016,Chen2018}. Further, since these methods bias the dynamic behavior of the system, estimation of kinetic properties is only possible under restrictive conditions \cite{Voter1997}. A complementary line of work seeks to increase the integration time-step directly, reducing the number of integration steps per unit time. Despite decades of intense research in this direction \cite{Tuckerman1992,Feenstra1999,Leimkuhler2013}, integration steps remain on the femtosecond scale, leaving even the most efficient simulations orders of magnitude too slow to capture experimentally relevant molecular processes.

A parallel line of progress has focused on harnessing ever-larger computational resources. Specialized compute architectures \cite{Shaw2021,Ohmura2014} have achieved continuous millisecond-scale trajectories for small proteins, revealing mechanistic detail inaccessible to conventional hardware \cite{Shaw2010}. Distributed platforms such as Folding@home leverage millions of short trajectories contributed by volunteers \cite{Shirts2000}, while modern GPU-based algorithms have brought comparable acceleration to widely used MD engines \cite{Eastman2023,Abraham2015,Harvey2009,Case2023}. Together with statistical frameworks such as Markov state models (MSMs) \cite{Swope2004,Buchete2008,Prinz2011,msmbook}, these efforts have enabled the reconstruction of long-timescale kinetics from massive ensembles of short simulations. Yet all remain bound by the need to generate femtosecond-resolved trajectories, keeping progress tied to extreme computational resources. A conceptually different approach would be to model the effective long-lag dynamics directly, without resorting to brute-force sampling or biasing \cite{Olsson2026}.

MD simulations function through the numerical integration of the Langevin equation \cite{Langevin}. As we move along the simulation trajectory, MD generates statistical samples from a transition probability distribution $p(\x_{t+\tau}\mid \x_t)$ where $\tau$ is on the order of femtoseconds, and $\x_{t}$ and $\x_{t+\tau}$ are points in the phase-space. Crucially, this distribution is not {\it ad hoc}: it approximates the Green’s function of the Fokker-Planck equation governing Langevin dynamics, providing the theoretical foundation for viewing MD trajectories as stochastic samples from an underlying probabilistic process \cite{Risken1996}. It follows that analogous transition probability distributions exist for much larger $\Delta t$, and that these can, in principle, be learned directly for a given molecular system \cite{ito}. Learning such long-lag transition densities offers a direct route to coarse-grained yet statistically faithful dynamics, sidestepping the need for explicit time integration.

Here, we introduce Transferable Implicit Transfer Operators (TITO), a deep generative framework that learns these transition probability distributions across molecular systems. TITO allows us to choose the simulation step size freely, whether to match the characteristic timescales of experiments or to accelerate sampling of slow conformational transitions. Trained on MD data from small molecules and short peptides, TITO simultaneously learns transitions at multiple step sizes, ensuring consistency with the underlying stochastic process. As a result, it preserves key statistical properties such as Boltzmann equilibrium, Markovianity, and relaxation dynamics, suggesting approximate energy conservation and equipartition.

TITO demonstrates quantitative transferability to molecular systems of similar size as in the training data, and provides qualitative insights for molecules twice as large. Unlike conventional simulation-based sampling, TITO offers explicit control over the trade-off between accuracy and computational cost, enabling speedups of up to 15,000-fold. By learning effective long-lag dynamics directly, TITO takes a step toward bridging the longstanding gap between atomistic resolution and experimentally relevant timescales. More broadly, it establishes a new paradigm for accelerating molecular simulations, with the potential to extend atomistic modeling to processes previously beyond reach.


\section*{Transferable Implicit Transfer Operators}
\begin{figure}
\centering
\includegraphics[width=\linewidth]{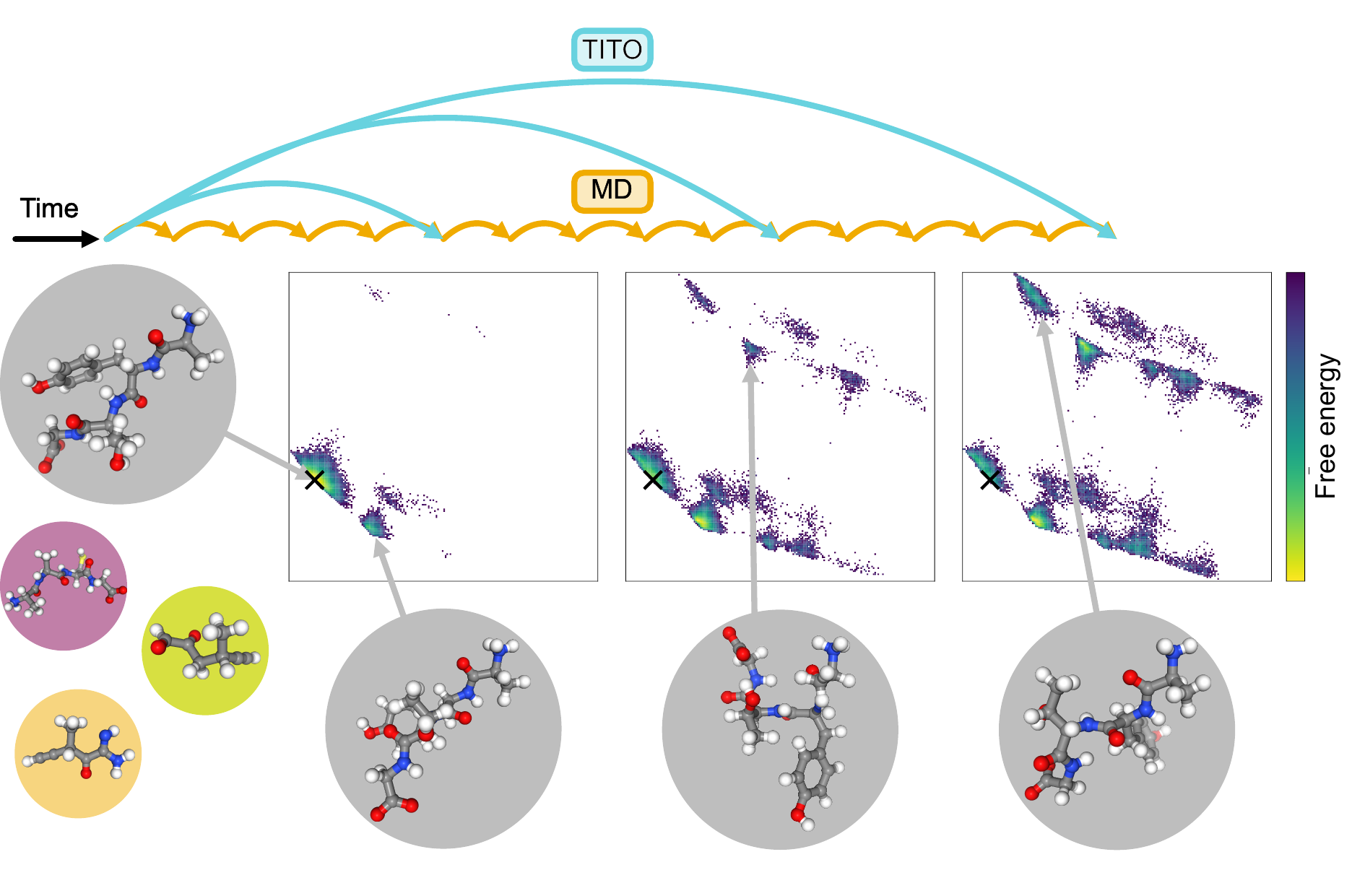}
\caption{\textbf{Transferable Implicit Transfer Operators (TITO)}: A multi–timescale surrogate model for molecular dynamics that is transferable across systems. Starting from an initial condition (black cross), TITO generates molecular dynamics ensembles for diverse molecules at arbitrary lag times.} 
\label{fig:method}
\end{figure}

At its core, TITO (Transferable Implicit Transfer Operators, Fig.~\ref{fig:method}) learns the effective rules of molecular motion: predicting how atomic configurations evolve over time without explicit time integration. Rather than advancing dynamics through numerous femtosecond scaled-steps, TITO draws statistical samples directly from the transition distribution $p(\mathbf{x}_{t+\Delta t}\mid \mathbf{x}_t)$, capturing how configurations change over a specified lag time $\Delta t$ which may be arbitrarily long. Trained across diverse molecular systems and lag times, TITO generalizes both across chemistry and temporal scale.

Training proceeds from reference molecular dynamics trajectories simulated with a small integration step \(\tau\):
\begin{equation}
\mathbf{X} = \{\mathbf{x}_\tau, \dots, \mathbf{x}_{N\tau}\}, \quad 
\mathbf{x}_{n\tau} \sim p(\mathbf{x}_{n\tau} \mid \mathbf{x}_{(n-1)\tau}), \quad n=1,\dots,N ,    
\end{equation}

collected across a diverse set of molecules. From these data, the model learns to reproduce the time-integrated transition statistics that would arise if the dynamics were propagated at much larger effective steps $\Delta t = m\tau$, where $m$ is an arbitrary large integer.

We parametrize the transition probability distribution, using a continuous normalizing flow (CNF) through the equivariant \emph{flow matching} \cite{flowmatching,equivariantflowmatching} objective. A CNF consists of an ordinary differential equation (ODE) and an easy-to-sample `base distribution,' $p_0$, such as a Gaussian \cite{1806.07366}. The velocity field of the ODE is parameterized with a neural network model which is trained to ensure that the resulting flow transports samples from $p_0$ to a distribution $p_1$ closely matching the target data distribution, here, the transition probability distribution. The flow is then the set of all integral paths \(\mathbf{x}_{t+\Delta t}^T\), where \(T \in [0,1]\) is the ODE integration time. Throughout this work, superscripts denote ODE integration time, while subscripts indicate MD simulation time.

In practice, we learn the weights $\theta$ of a neural network, $\mathbf{v}_\theta \left( \x_{t+\Delta t}^T; \ \x_t, \ \Delta t, \ T \right)$, to match a conditional flow which approximates the transition probability distribution, by minimizing the conditional flow matching loss,
\begin{equation}
\mathcal{L}(\theta) = 
\mathbb{E}_{\x_{t},\x_{t+\Delta t} \sim \mathbf{X},\, T \sim \mathcal{U}(0,1)} 
\Big[ \big\| \mathbf{v}_\theta \left( \x_{t+\Delta t}^T; \ \x_t, \ \Delta t, \ T \right) - (\x_{t+\Delta t}^1 - \x_{t+\Delta t}^0) \big\|^2 \Big].    
\end{equation}

During training, we sample molecules and lag times jointly, enabling TITO to generalize across both chemical composition and temporal scale. After training, new trajectories are generated by sampling from \(p_0\) and integrating the learned ODE defined by \(\mathbf{v}_\theta\). Full model details are provided in Section~\nameref{sec:model}.

We train TITO models on two datasets. The first, \textbf{MDQM9-nc} \cite{sma-md}, contains MD simulations of 
small organic molecules, while the second, \textbf{Timewarp} \cite{timewarp}, provides tetra-peptides trajectories. Together, these datasets enable training across a range of molecular sizes and chemistries. We provide details on dataset generation and pre-processing in Section~\nameref{sec:data}.

\section*{Results}


\subsection*{Integrity of the Boltzmann distribution under TITO dynamics in unseen small molecules and peptides}

\begin{figure}
\vspace{-2 cm}
\centering
\includegraphics[width=0.9
\linewidth]
{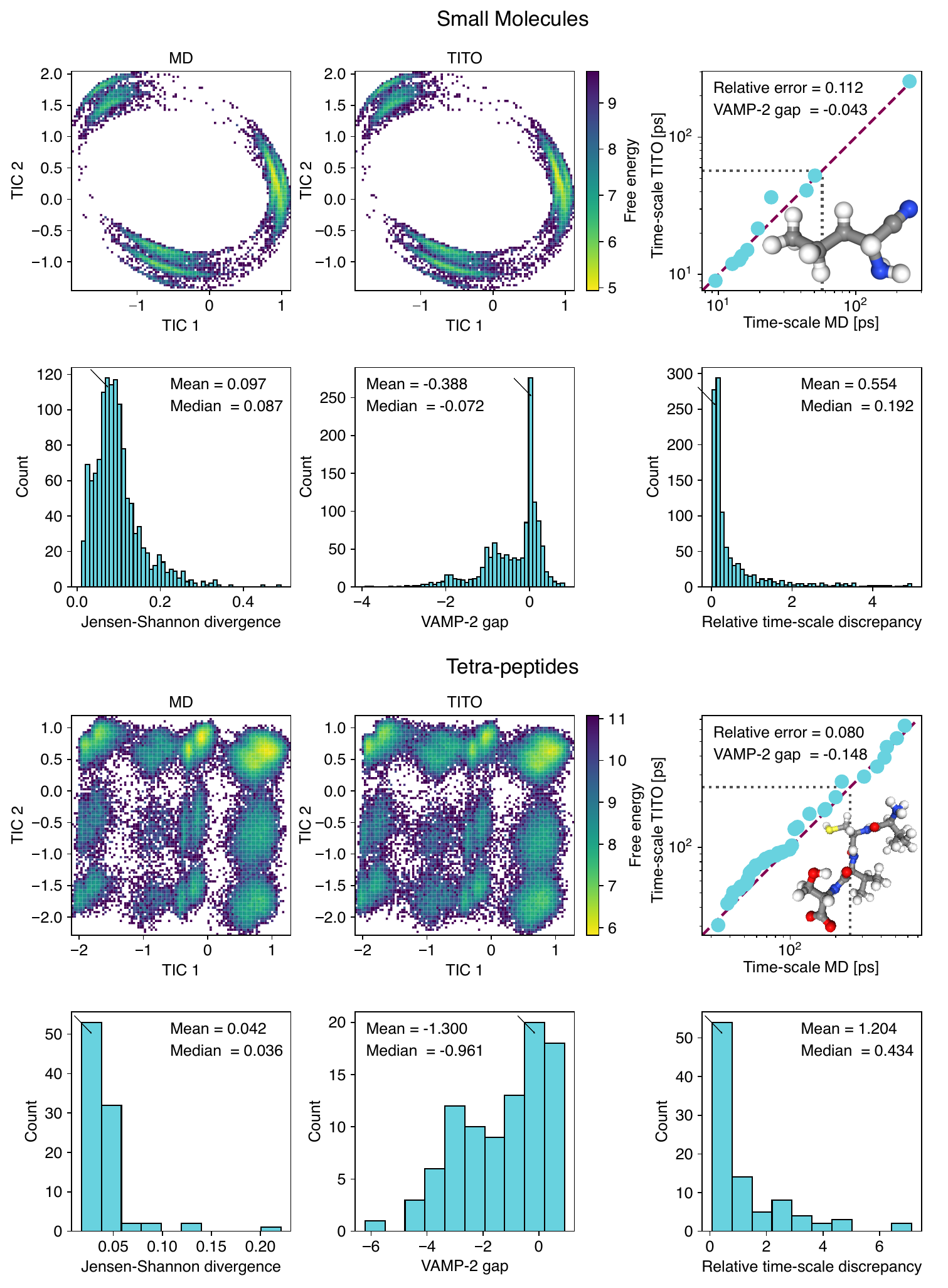}
\caption{{\bf TITO accurately predicts both thermodynamics and kinetics.} Small molecules (top) and tetra-peptides (bottom).
Top row: Projection onto the first two TICs (time-lagged independent components) and comparison of VAMP timescales between MD and TITO-generated samples for a representative molecule.
Bottom row: Aggregated evaluation across systems: Jensen–Shannon divergence in TIC subspace (left), VAMP-2 gap (center), and top-10 relative error (right). Black arrows denote the position of the example molecule within each histogram.} 
\label{fig:mdqm9_tw}
\end{figure}

A defining property of molecular systems undergoing Langevin dynamics is convergence to the Boltzmann distribution. In contrast, when a generative model is trained to approximate time–integrated transition probabilities, this guarantee is no longer automatic. The central question is therefore whether TITO preserves physical realism---whether it samples configurations consistent with the Boltzmann distribution---or instead produces unphysical states, analogous to large language models generating text that is fluent but factually incorrect. 

To test this, we examined whether the Boltzmann distribution, $\mu\propto\exp(-\beta U(\x))$, of the potential energy function, $U$, at inverse temperature $\beta$ is the invariant measure (i.e., stationary distribution) of the transfer operator implicitly learned by TITO \cite{bopito}. Because the learned operator is not directly accessible, we assessed this property numerically. Specifically, we drew initial conditions $\x_0 \sim \mu$ from long unbiased MD simulations, generated trajectories using $p(\x_{\Delta t} \mid \x_0)$ with TITO, and compared the empirical distributions of $\x_0$ and $\x_{\Delta t}$ using the Jensen–Shannon divergence (JSD) \cite{Lin1991}. Across a broad set of small molecules and tetrapeptides unseen during training, TITO reproduced the Boltzmann distribution obtained from reference MD simulations (Fig.~\ref{fig:mdqm9_tw}). A small fraction of cases exhibited elevated JSD values, indicating discrepancies that could reflect either spurious (hallucinatory) samples or genuine metastable states not explored by reference simulations (Suppl.~Fig.~\ref{fig:si-examples-jsds}).

To probe these outliers, we constructed Koopman operator models from long unbiased MD and from TITO trajectories. These models characterize the system’s slowest dynamical modes and associated metastable states. We found that TITO generally reproduced relaxation timescales of MD, suggesting that much of the JSD tail arises from minor numerical mismatches. However, a subset of systems displayed substantially slower relaxation times under TITO, as revealed by significantly larger VAMP (Variational Approach to Markov Processes) scores \cite{VAMP}. Because theoretical results show that timescales are bounded from above \cite{Nske2014}, this indicates that TITO sampled metastable states not observed in the MD trajectories.

\begin{figure}
\centering
\includegraphics[width=\linewidth]{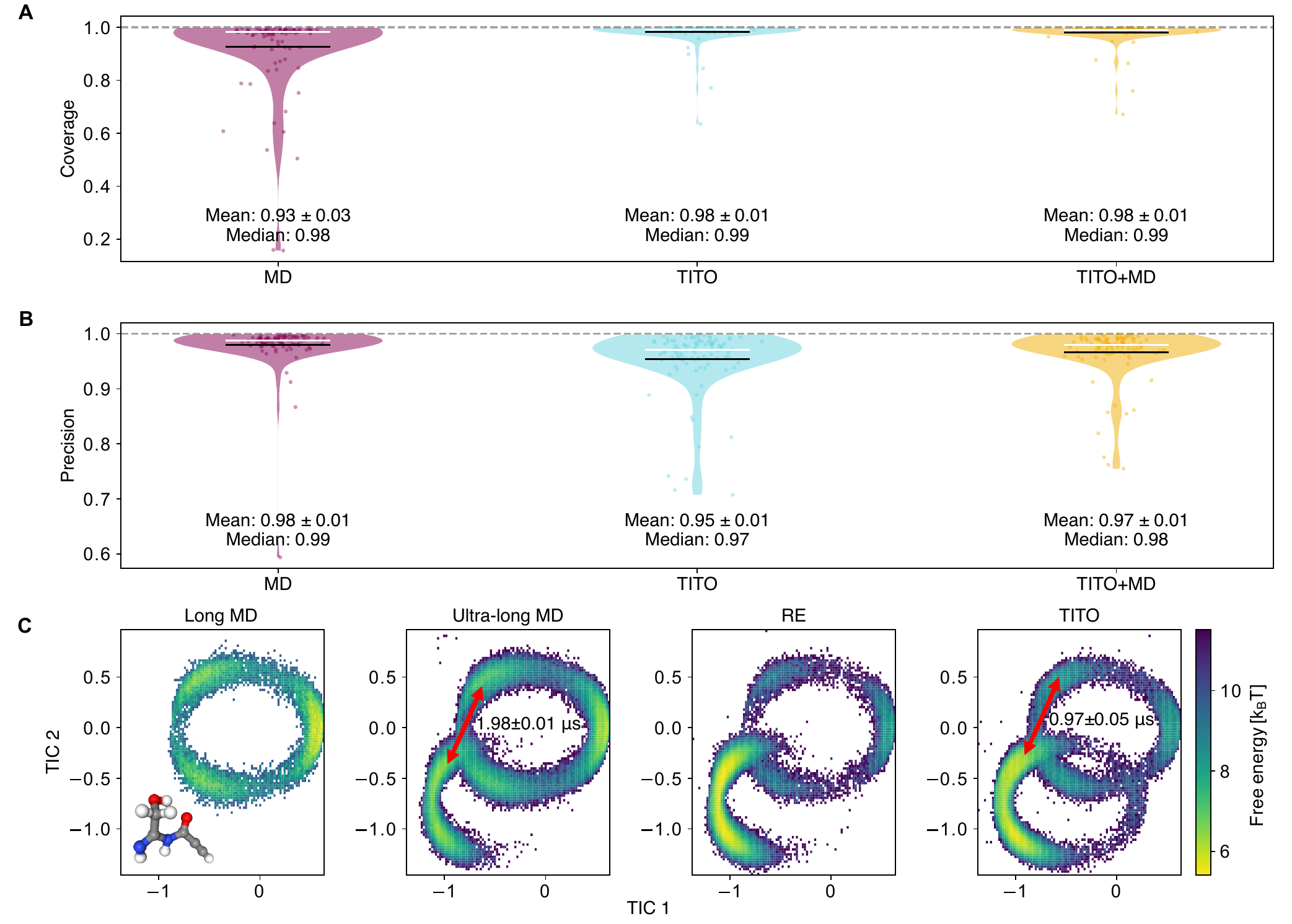}
\caption{{\bf TITO accurately samples two orders of magnitude slower dynamics than the training data.} (A) Coverage and (B) precision, see \nameref{sec:metrics} section, w.r.t to RE of projections onto the first two TICs for training set-like simulations (MD, 36.5 $\mathrm{ns}$), TITO (32 $\mu \mathrm{s}$) and an ensemble of ultra-short MD initialized with TITO samples (10 $\mathrm{ps}$ per sample). TITO achieves better coverage of the conformational space than MD. (C) Example propiolamide test-set molecule: Long MD (36.5 $\mathrm{ns}$) fails to sample the most meta-stable basin, while TITO (160 $\mu\mathrm{s}$, initialized from long MD) successfully transitions into the dominant basin, recovers the Boltzmann distribution and estimates the right order of magnitude for the corresponding VAMP implied timescale, estimated from ultra-long MD (16 $\mathrm{ms}$). Structural insights of this transition and alternative TICA projections are provided in Suppl.~Figs.~\ref{fig:si-mol21-torsions} and ~\ref{fig:si-long-tica}, respectively.} 
\label{fig:long-term}
\end{figure}

To evaluate whether these new states were physically meaningful, we performed extensive replica exchange (RE) MD simulations \cite{re}. On average, TITO covered all density regions visited by RE MD, whereas long conventional MD failed to do so in a significant fraction of systems (Fig.~\ref{fig:long-term}A). States detected by TITO but absent in long MD were consistently recovered in RE MD (Suppl.~Fig.~\ref{fig:si-tito-better-md}), confirming that they correspond to genuine metastable basins. In one representative example, a propiolamide, TITO uncovered a metastable basin absent from long MD but corroborated by both ultra-long unbiased MD and RE MD simulations (Fig.~\ref{fig:long-term}C  and Suppl.~Figs.~\ref{fig:si-mol21-torsions} and~\ref{fig:si-long-tica}). This metastable basin was not sampled when the model was trained on a training-set–length trajectory of this molecule (Suppl. Fig.~\ref{fig:si-one-mol-state-discovery}), showing that it emerges from multi-molecule training. Remarkably, although trained only on nanosecond MD data, TITO correctly inferred an exchange timescale between basins on the order of microseconds---consistent with estimates from ultra-long trajectories.

Intriguingly, TITO also predicted additional states not observed in either reference method (Fig.~\ref{fig:long-term}B). To further investigate these cases, we initialized ensembles of nanosecond-length unbiased MD trajectories from TITO-generated configurations. These simulations exhibited a small but systematic improvement in agreement with RE MD (Suppl.~Fig.~\ref{fig:si-js-vs-mdft}), indicating that TITO samples near-physical configurations capable of relaxing into correct basins under explicit dynamics. We then assessed the stability of these newly identified states in ensemble simulations and found that the configurations remained metastable (Suppl. Fig.~\ref{fig:si-meta-stable}), suggesting that they represent physically valid states rather than artifacts of the learned dynamics. Collectively, these findings demonstrate that TITO not only preserves the integrity of the Boltzmann distribution but also uncovers metastable states that would likely remain undetected using conventional methods within practical computational limits.

We further examined whether TITO’s generalization correlates with chemical similarity between training and test molecules. Surprisingly, no such correlation was observed in either of the data-sets (Suppl.~Figs.~\ref{fig:si-similarity-mdqm9} and \ref{fig:si-similarity-tw} and Suppl.~Table \ref{table:si-similarity-tw}). This lack of correlation suggests that chemical composition alone provides limited signal for guiding iterative refinement or active learning of generative dynamical models, underscoring the need for alternative strategies to improve generalization.

\subsection*{TITO faithfully reproduces relaxation transients in unseen molecular systems}
Next, we investigate whether the dynamics generated by TITO is statistically equivalent to that generated by numerical MD simulations. Since we here target MD in the NVT ensemble--- keeping the number of particles (N), volume (V), and temperature (T) constant during a simulation---the dynamics are stochastic, and consequently, we use statistical tools to quantitatively compare the two approaches.

A stringent test of dynamical fidelity is whether a model can reproduce relaxation processes across systems that differ vastly in their intrinsic timescales. As noted TITO accurately recapitulates relaxation dynamics in molecules and peptides outside its training set, and crucially provides quantitative predictions spanning over three orders of magnitude in characteristic times (Suppl.~Fig.~\ref{fig:si-slow-and-fast}). Predicted timescales remain faithful w.r.t. MD and consistent with the variational principle for Markov processes \cite{Nske2014} across different lag times (Suppl.~Fig.~\ref{fig:si-ts-consistency}). This level of agreement indicates that TITO has learned an effective and generalizable representation of the underlying stochastic dynamics rather than merely fitting short-time correlations. 

\begin{figure}
\includegraphics[width=1\linewidth]{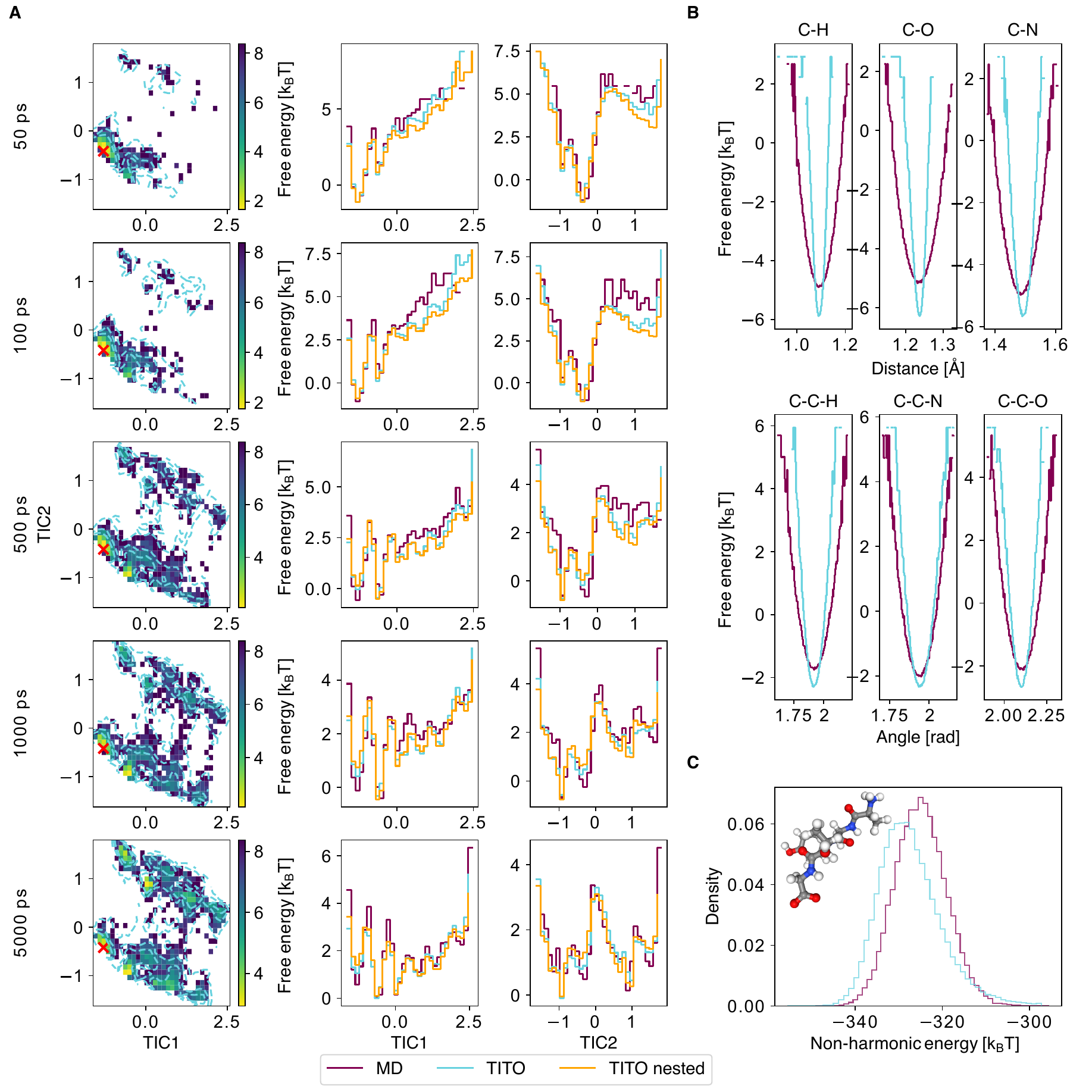}
\caption{{\bf TITO recapitulates transients, fast vibrations and potential energies. }(\textbf{A}) Free energy of the transition probability estimated with TITO for AYTG (test set) at increasing lag times (top to bottom). The first column shows conditional free energies projected onto the first two TICA components (TICs). Contours represent TITO samples, while 2D histograms correspond to Markov state model estimates from MD data. The red cross marks the simulation’s initial state. The second and third columns display marginal distributions along each TIC. Nested samples are generated with 5 steps. (\textbf{B}) Free energy profiles of bond distances (top) and angles (bottom). (\textbf{C}) Probability density of the non-harmonic potential energy.} 
\label{fig:marginals}
\end{figure}

Still, matching timescales alone does not guarantee that the associated motions are physically meaningful. To examine this, we performed extensive TITO simulations and compared the full relaxation transients of slow dynamical modes with those obtained from long, unbiased MD trajectories. Across two orders of magnitude in timescale, the agreement was striking, suggesting that TITO reproduces both the rate and the mechanism of the underlying molecular dynamics (Fig.~\ref{fig:marginals}A). This demonstrates that the learned transition operator generalizes dynamically, faithfully capturing the hierarchy of molecular motions that govern relaxation kinetics. 

A further requirement is internal consistency across time resolutions. Because TITO predicts time-integrated dynamics at multiple timescales, the resulting transients must be consistent regardless of whether they are generated in a single long step or as a sequence of shorter steps (nested sampling). We find that the relaxation transients remain self-consistent under this test (Fig.~\ref{fig:marginals}A), suggesting that the learned transition density satisfies the Chapman–Kolmogorov equation and thus encodes genuinely Markovian dynamics. We remark that even if Chapman-Kolmogorov consistency is not hard-coded into the model architecture, it is implicitly promoted by the Implicit Transfer Operator objective. Because the model is trained to learn conditional distributions at multiple lag times, the loss encourages coherence across compositions of the learned dynamics.

Finally, we investigated whether the high fidelity of slow dynamics is achieved at the expense of accuracy in fast, rapidly relaxing modes. Remarkably, despite operating at time steps orders of magnitude larger than those of bond and angle vibrations, TITO accurately reproduced equilibrium properties (Fig.~\ref{fig:marginals}B) and generated conformations with potential energies closely matching those from unbiased MD (Fig.~\ref{fig:marginals}C). The main deviation we observed was a slight underestimation of the variance of fast modes, which in turn leads to systematically lower potential energies relative to reference simulations.




\subsection*{Qualitative extrapolation to larger peptide systems}
We next examined whether TITO can extrapolate beyond the molecular sizes represented in its training data. This setting provides a stringent test of transferability: a model trained on short peptides must infer effective dynamics at new length scales, where both the number of atoms and the hierarchy of internal motions increase substantially. Specifically, we applied a model trained only on tetrapeptides to generate trajectories for penta-, hexa-, hepta-, and octapeptides.

Extrapolation introduces a scale mismatch in the latent base distribution $p_0$, whose variance depends on system size. To mitigate this, we rescaled the standard deviation of $p_0$ according to Flory’s scaling law for the radius of gyration of random polymers, $\langle R_g \rangle \propto N^{0.688}$ \cite{flory1953principles}, where $N$ denotes the number of residues. Guided by this simple physical prior, TITO produced configurations with realistic local geometry and global compactness across peptide lengths; without this correction, stable extrapolation beyond pentapeptides was not achievable.

\begin{figure}
\centering
\includegraphics[width=\linewidth]{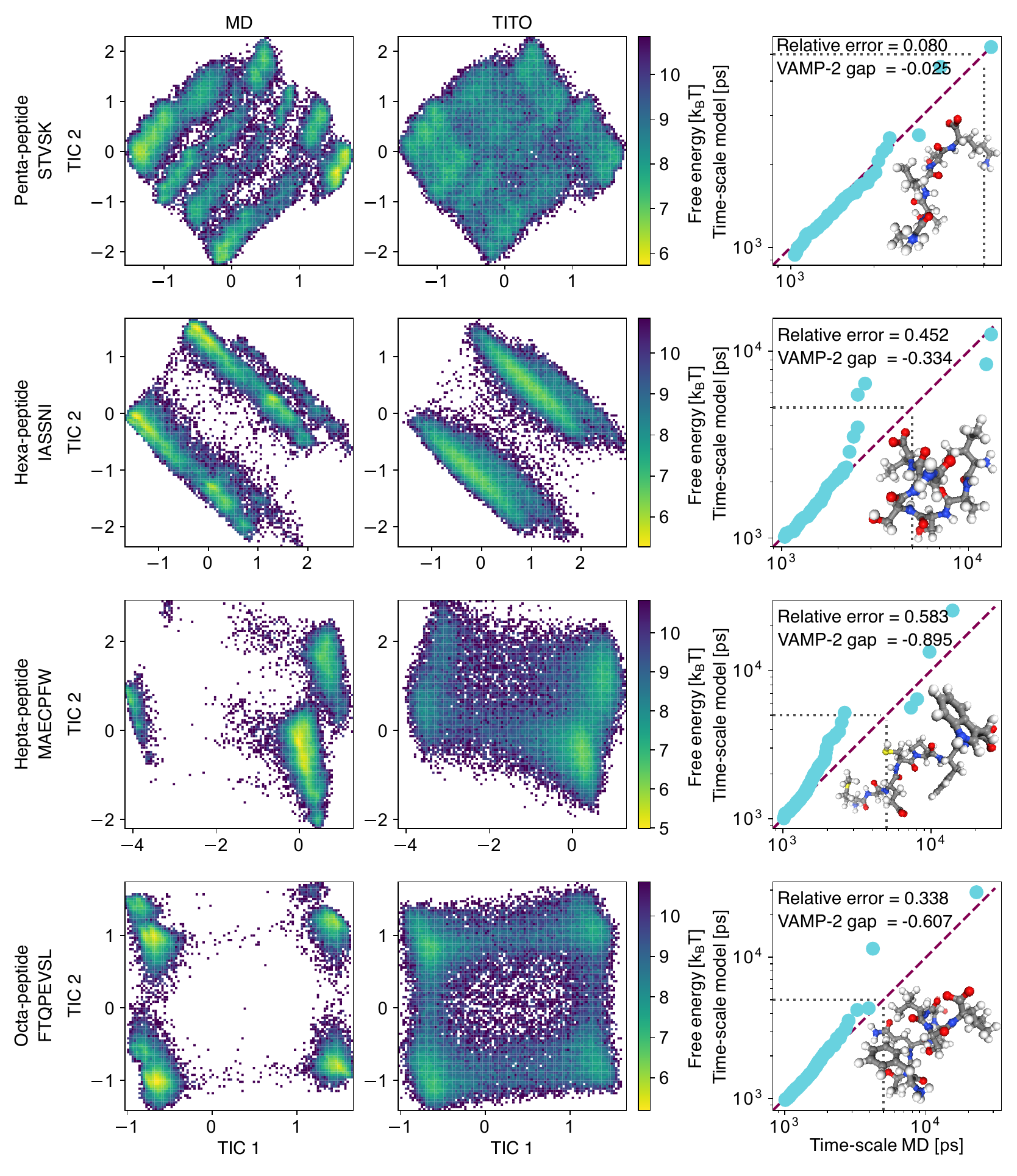}
\caption{{\bf Extrapolation to larger systems.} Free energy landscape and VAMP timescales of a TITO model trained on tetra-peptides and performing 5 $\mathrm{ns}$ single-step sampling for penta, hexa and hepta and octa- peptides. } 
\label{fig:extrapolation}
\end{figure}

With the scaling correction in place, TITO approximately recovered the conformational landscapes of larger peptides and reproduced relaxation times qualitatively consistent with explicit MD, even when the sequence length was doubled relative to the training systems (Fig.~\ref{fig:extrapolation}). For the largest peptides, however, the generated trajectories exhibited mild structural compaction (Suppl.~Fig.~\ref{fig:si-rg-extra}) and a systematic downward drift in potential energy (Suppl.~Fig.~\ref{fig:si-energy-extra}), leading to instability in long nested-sampling runs. These deviations likely arise from cumulative local errors that are amplified at increasing system sizes. Nevertheless, the generated configurations remain physically meaningful and can be readily refined by short low-cost MD equilibrations.

Together, these results demonstrate that TITO captures transferable physical principles sufficient to generalize far beyond its training domain, while also delineating the limits of such extrapolation. The observed degradation at large sizes naturally motivates hybrid divide-and-conquer strategies, in which long TITO propagation steps are interleaved with brief MD equilibration phases---akin to hybrid Monte Carlo or multi-resolution simulation schemes \cite{Nilmeier2011,Ross2020,sma-md}.

\subsection*{Calibration of simulation accuracy to compute budget}
For practical impact, TITO must deliver substantially higher throughput than conventional MD while retaining quantitative accuracy in equilibrium and dynamical properties. Two factors determine the effective simulation throughput: (i) the molecular size, which constrains the number of simulations that can be run in parallel on a GPU with fixed memory, and (ii) the number of ODE solver steps required for each CNF evaluation, which controls the cost per TITO step.

We find that equilibrium properties can be reproduced at comparatively low computational budgets, whereas accurate estimation of relaxation timescales requires additional solver steps and hence higher cost (Suppl.~Fig. \ref{fig:si-compute-calibration}). This trade-off implies that the compute budget can be calibrated to match the target application, for example, prioritizing thermodynamic ensemble generation versus reproducing kinetic observables measured in experiment.

To quantify achievable throughput, we report the maximum simulation time reached on a single GPU. As shown in Table~\ref{table:simulation_time}, TITO enables approximately 10 milliseconds of physical simulation time per day of computation, representing a four-order-of-magnitude improvement relative to standard unbiased MD simulations using the same resources. These results suggests that TITO can be tuned flexibly: users may trade simulation fidelity against throughput depending on the level of accuracy required. Further, we emphasize that these gains can potentially be even larger if the model is trained to predict larger time steps, paving the way to study ultra-slow processes in biology and and material science.

\begin{table}[]
\centering
\begin{tabular}{|c|c|c|}
\hline
Systems                                         & Method & Simulation throughput per day in one GPU \\ \hline
\multirow{2}{*}{Small molecules}                          & MD     &  3.5 $\mu\mathrm{s}$                     \\ \cline{2-3} 
                                & TITO   & 11.3 $\mathrm{ms}$ \\ \hline
\multicolumn{1}{|l|}{\multirow{2}{*}{Tetra-peptides}} & MD     & 0.67   $\mu \mathrm{s}$                  \\ \cline{2-3} 
\multicolumn{1}{|l|}{}                          & TITO   & 10.3 $\mathrm{ms}$                     \\ \hline
\end{tabular}
\caption{{\bf Maximum simulation time throughput.} Estimates using one NVIDIA A100 80 GB GPU for one day.}
\label{table:simulation_time}
\end{table}

\section*{Discussion}

We introduce Transferable Implicit Transfer Operators (TITO), a chemically generalizable generative model that reproduces molecular dynamics at a fraction of the computational cost of traditional simulations. TITO quantitatively recovers both the equilibrium probabilities of molecular configurations and the rates and mechanisms of conformational exchange across diverse chemistries, from small molecules to peptides. In essence, it delivers the fidelity near that of molecular dynamics at the cost of sampling a deep generative model.

TITO achieves this by learning the statistics of time-integrated dynamics directly from simulation data, allowing propagation over arbitrarily long lag times without explicit numerical integration at femtosecond resolution. This formulation yields an acceleration of up to four orders of magnitude in the quantitative characterization of equilibrium states and relaxation kinetics at  compute cost parity. Furthermore, TITO retains predictive power beyond its training regime, qualitatively reproducing thermodynamic and kinetic behavior in peptides up to twice the size of those used for training highlighting its ability to extrapolate across molecular size.

TITO differs fundamentally from dominant paradigms in generative models of molecular dynamics, Boltzmann Generators which aim to quantitatively sample the independent equilibrium samples from the Boltzmann distribution \cite{bgs} and Boltzmann Emulators which sacrifice quantitative alignment with MD to boost efficiency and scaling \cite{bowenjing_torsional, sma-md, bioemu}. These methods, and in particular their transferable variants \cite{transbgs,bioemu,sma-md}, are rapidly becoming a viable complement to MD simulations when equilibrium properties are the target of investigation. However, these approaches cannot capture dynamic properties such as rates or mechanisms.

Other methods that aim to predict the dynamics provide mainly qualitative insights and at fixed timescales \cite{timewarp, mdgen,2503.23794, costa2024equijumpproteindynamicssimulation, murtada2025mdllm1largelanguagemodel} or enable much smaller time steps \cite{bigi2025flashmdlongstrideuniversalprediction, Vlachas2021}. Recent work explored multi–timescale training in small proteins \cite{costa2025acceleratingproteinmoleculardynamics}. To our knowledge, TITO is the first framework to achieve physically realistic, multi–timescale sampling with demonstrated transferability across both chemical composition and molecular size. Other complementary strategies, include machine-learning infused path-sampling \cite{Jung2023} strategies or latent space simulators \cite{Sidky2020,Vlachas2021, Wang2024} show promise in scaling to larger systems, but generalization remains an open challenge requiring careful modeling and calibration for every specific process of interest.

Machine learned interatomic potentials\cite{Kabylda2025, Batatia2025} and coarse-grained force-fields \cite{Charron2025, Majewski2023} have similarly shown impressive strides towards general purpose transferability. These models can guarantee realistic physical dynamics, depending on the integration strategy chosen. So while they might boost the accuracy over current force-fields and coarse-grained models, they still rely on iterative numerical integration with tiny time steps making their computational footprint significant. TITO instead offers a paradigm shift: bypassing iterative integration altogether.

Despite these advances, important limitations remain. At present, TITO is restricted to implicit solvent representations and system sizes of at most a few hundred atoms. Extending the method to explicitly solvated biomolecules with tens to hundreds of thousands of degrees of freedom will require innovations in neural architectures \cite{euclideanfastattention, semlaflow, 2502.13797,NEURIPS2024_fad8e191}, such as equivariant transformers \cite{fuchs2020se3transformers3drototranslationequivariant}, linear attention \cite{katharopoulos2020transformersrnnsfastautoregressive} or Euclidean fast attention \cite{euclideanfastattention}, and/or hierarchical strategies such as coarse-graining \cite{ml-coarse-grain}. In addition, periodic boundary conditions—essential for realistic modeling of solvated systems---are not yet supported. While our experiments show promising extrapolation to larger and chemically distinct molecules, generalization performance still depends on the chemical similarity between target and training systems. Achieving broad chemical coverage will necessitate larger and more diverse training datasets. Finally, TITO is presently limited to a single thermodynamic state (NVT ensemble at room temperature). Extending thermodynamic transferability, including across temperatures, pressures or even force fields, would enable the study of the influence of thermodynamic perturbations on molecular systems stationary and dynamic properties. Promising prior work has demonstrated that thermodynamic transferability can be achieved by additional conditioning variables through learned embeddings of the thermodynamic state \cite{thermodynamicinterpolation,Herron2024}, which motivates its applicability to kinetic modeling as well..

Intriguingly, we find that TITO's generalization performance shows no clear relationship to the chemical similarity between training and test systems. This observation challenges the prevailing assumption that broader chemical coverage alone ensures generalization. Instead, it suggests that the structure and diversity of training data---how well they represent relevant dynamical motifs and energy landscapes---may be more critical than sheer data volume. In this view, progress may hinge less on scaling to ever-larger simulation datasets and more on carefully curated, mechanistically diverse benchmarks that capture the essential physics of molecular dynamics. 

In summary, TITO establishes a new paradigm for transferable generative modeling of molecular dynamics, unifying thermodynamic sampling and dynamical prediction in a singular generative surrogate. By enabling accelerated and chemically transferable estimation of stationary and dynamic properties---such as free energies and rates---TITO paves the way toward practical deep-learning–based acceleration of molecular simulations.






\section*{Materials and methods}
\subsection*{Data} 
\phantomsection
\label{sec:data}
We use three datasets covering different regions of chemical space:
\begin{enumerate}
    \item \textbf{Small molecules}: The MDQM9-nc dataset \cite{sma-md}  contains MD simulations for 12,530 small non-cyclic molecules from the QM9 dataset \cite{rupp}. Simulations are performed in a vacuum, at room temperature using the GAFF force field \cite{Wang2004}. Simulation time is dependent on the molecule size  with a median sampling time of 36.5 $\mathrm{ns}$. We perform extra 1 $\mu\mathrm{s}$ RE simulations across 8 temperatures (300, 400, 500, 600, 700, 800, 900 and 1000 K). The average exchange rate 58 \%.
    \item \textbf{Tetra-peptides}: The Timewarp dataset\cite{timewarp} contains MD simulations for tetra-peptides. It contains two tetra-peptides sub-datasets, large, which contains train 1457 molecules and huge, with 92 larger molecules . We use large as training data and huge as test set. The simulations are performed in implicit water and at room temperature. Simulation time is 50 $ns$ for training set molecules and 1 $\mu\mathrm{s}$ for test set molecules.
    \item \textbf{Larger systems}: We performed 1 $\mu\mathrm{s}$ simulation of penta-, hexa-, hepta- and octa-peptides with the same simulation parameters as in the Timewarp dataset. For each peptide length, six sequences were randomly sampled based on vertebrate amino acid frequencies.
\end{enumerate}

\subsection*{Model}
\phantomsection
\label{sec:model}
TITO uses equivariant optimal transport flow matching to parameterize the transition probability. Flow matching~\cite{flowmatching} provides an efficient framework for training continuous normalizing flows by aligning a learnable velocity field with the optimal transport velocity field between an easy to sample base distribution $p_0$ and a target distribution $p_1$. Rather than directly minimizing a divergence between the generated distribution and $p_1$, flow matching constructs intermediate states along an interpolation between $p_0$ and $p_1$. The neural ODE vector field is then trained to predict the conditional displacement between these paired states. We use linear interpolants \cite{JMLR:v26:23-1605},
\begin{equation}
\x^T = (1-T)\,\x^0 + T\x^1, \quad T \in [0,1],    
\end{equation}
with target velocity,
\begin{equation}
\mathbf{v}^T = \frac{d \mathbf{x}^T}{dT} = \mathbf{x}^1 - \mathbf{x}^0 .
\end{equation}

Molecular distributions live in a space with inherent symmetries, such as rotational and permutational invariances of atomic coordinates. Equivariant optimal transport \cite{equivariantflowmatching} incorporates these symmetries to construct shorter paths by aligning sample pairs, $\x^0$ and $x^1$,  along their symmetry orbits. In practice, this minimization is approximated sequentially: the optimal permutation by solving a linear sum assignment problem \cite{Crouse2016}, followed by an optimal superpositioning through solving a Procrustes problem \cite{Kabsch1976}. 

For the velocity field model,  \(\mathbf{v}_\theta\), we use a modified SE3-ITO architecture \cite{ito} enriched with edge features encoding interaction types between atoms. Specifically, we distinguish single, double, triple, and through-space bonds or interactions. As in SE3-ITO, we assume a complete interaction graph where every atom interacts with all others, with bonded interactions prioritized according to the order listed above.

Model training and inference parameters for different experiments are included in Suppl.~Tables \ref{table:si-train-params} and \ref{table:si-sampling-params}, respectively.

\subsection*{Evaluation metrics}
\phantomsection
\label{sec:metrics}

\paragraph{Jensen-Shannon Divergence (JSD)}
The Jensen-Shannon Divergence provides a symmetric measure of similarity between two probability distributions. Given two distributions $p$ and $q$, the JSD is defined as  
\begin{equation}
   \mathrm{JSD}(p \parallel q) = \frac{1}{2} D_{\mathrm{KL}}(p \parallel m) + \frac{1}{2} D_{\mathrm{KL}}(q \parallel m), 
\end{equation}

where $m = \tfrac{1}{2}(p + q)$ and $D_{\mathrm{KL}}$ denotes the Kullback--Leibler divergence. Unlike the KL divergence, the JSD is bounded between 0 and 1.

\paragraph{Free energy} The free energy (negative log-likelihood) of a sub-space of the conformational space $\Omega_i \in \Omega$ is 
\begin{equation}
F(\Omega_i) = - k_BT \ \log(p(\Omega_i)) \text{ with } p(\Omega_i) =  \int_{\Omega_i} p(\x)\, \mathrm d \x.    
\end{equation}
\paragraph{Coverage and precision.} 
Coverage and precision quantify the degree of overlap between the probability distributions generated by two sampling methods. 
Given two methods, $m_1$ and $m_2$, the coverage of $m_1$ with respect to $m_2$ is defined as
\begin{equation}
      \text{COV}_{m_1,m_2} = \int_{\Omega_{p_{m_1} \cap p_{m_2}}} p_{m_2}(\mathbf{x})\, \mathrm{d}\mathbf{x},  
\end{equation}

where $p_{m_i}$ denotes the probability mass sampled from method $m_i$, and 
$\Omega_{p_{m_1} \cap p_{m_2}} = \{\mathbf{x} : p_{m_1}(\mathbf{x}) > \delta \text{ and } p_{m_2}(\mathbf{x}) > \delta\}$. 
Precision is defined analogously as
\begin{equation}
    \text{PRE}_{m_1,m_2} = \int_{\Omega_{p_{m_1} \cap p_{m_2}}} p_{m_1}(\mathbf{x})\, \mathrm{d}\mathbf{x}.    
\end{equation}

Intuitively, coverage measures how much of the probability mass of $m_2$ is captured by $m_1$, while precision measures the fraction of $m_1$’s probability mass supported by $m_2$. 
In practice, we estimate these quantities from empirical histograms, defining $\Omega_{p_{m_1} \cap p_{m_2}}$ as the set of discrete states where both methods have nonzero counts.

\paragraph{Variational Approach for Markov Processes (VAMP)}
The VAMP framework provides a principled method for evaluating the quality of dynamical models based on the variational principle of conformation dynamics. We employed the following VAMP-based metrics:  

\paragraph{Implied timescales}  
The eigenvalues of the estimated transfer operator were used to compute implied timescales, defined as
\begin{equation}
t_i = -\frac{\tau}{\ln \sigma_i},    
\end{equation}
where $\sigma_i$ is the $i$-th singular value  of the Koopman operator approximation and $\tau$ the lag time.

\paragraph{Relative timescale discrepancy}  We define the relative timescale discrepancy as
\begin{equation}
 \epsilon_{rel} = \frac{1}{N} \sum_i^N \frac{\vert t_i^{MD} - t_i^{TITO} \vert}{t_i^{MD}},    
\end{equation}
where $t_i^{MD}$ and $t_i^{TITO}$ are implied timescales predicted with MD and TITO respectively and are sorted in decreasing order. We use  $N=10$ implied timescales throughout this work.
\paragraph{VAMP-2 score}  
The VAMP-2 score is the squared Frobenius norm of the singular value spectrum,
\begin{equation}
     \text{VAMP-}2^{(k)} = \sum_{i=1}^{k} \sigma_i^2.
\end{equation}   
Higher scores indicate that the model captures slow dynamical modes.  

\paragraph{VAMP-gap}  We define the VAMP-gap as the difference in VAMP2-scores between TITO and the MD,
\begin{equation}
\text{VAMP-gap} = \text{VAMP2-score}_\text{{MD}} - \text{VAMP2-score}_{\text{TITO}}.    
\end{equation}
Negative VAMP-gaps indicate TITO predicts slower dynamics and vice versa.


Together, these evaluation metrics provide complementary insights: the Jensen-Shannon Divergence and free energy excess measures how well the model reproduces equilibrium distributions, while VAMP metrics assess the model’s fidelity in capturing slow dynamical processes and metastability.

\section*{Acknowledgements}

\subsubsection*{Funding statement}
This work was partially supported by the Wallenberg AI, Autonomous Systems and Software Program (WASP) funded by the Knut and Alice Wallenberg Foundation (WASP AI-MLX Professorship to S.O.). Model training and inference was made possible by an allocation on the Berzelius resource provided by the Knut and Alice Wallenberg Foundation at the National Supercomputer Centre hosted by the National Academic Infrastructure for Supercomputing in Sweden (NAISS) (project: Berzelius-2025-189, to S.O.), partially funded by the Swedish Research Council through grant agreement no. 2022-06725. S.O. acknowledges funding from the Chalmers Academic Excellence Program.
\pagebreak
\subsubsection*{Author contributions}
\begin{table}[h!]
    \centering
    \renewcommand{\arraystretch}{1.5} 
    \begin{tabular}{l p{12cm}}
        \hline
        \textbf{Author} & \textbf{CRediT} \\
        \hline
        Simon Olsson & Writing - original draft, Conceptualization, Writing - review \& editing, Methodology, Resources, Funding acquisition, Data curation, Validation, Supervision, Software, Project administration, Visualization \\
        Mathias Schreiner & Conceptualization, Investigation, Methodology, Data curation, Validation, Formal analysis, Software \\
        Juan Viguera Diez & Writing - original draft, Conceptualization, Investigation, Writing - review \& editing, Methodology, Funding acquisition, Data curation, Validation, Supervision, Formal analysis, Software, Project administration, Visualization \\
        \hline
    \end{tabular}
\end{table}
\subsubsection*{Competing Interests Statement}
All authors declare that they have no competing interests.

\subsubsection*{Data, Code, and Materials Availability Statement}
All data needed to evaluate and reproduce the results in the paper are present in the paper and/or the Supplementary Materials. Code is available on the public repository \url{https://github.com/olsson-group/tito}. The repository also includes instructions to retrieve data used to train the model, and model weights used to generate the figures presented in the manuscript. This study did not generate new materials.

\bibliography{scibib}

@article{bgs,
  title = {Boltzmann generators: Sampling equilibrium states of many-body systems with deep learning},
  volume = {365},
  ISSN = {1095-9203},
  url = {http://dx.doi.org/10.1126/science.aaw1147},
  DOI = {10.1126/science.aaw1147},
  number = {6457},
  journal = {Science},
  publisher = {American Association for the Advancement of Science (AAAS)},
  author = {Noé,  Frank and Olsson,  Simon and K\"{o}hler,  Jonas and Wu,  Hao},
  year = {2019},
  month = sep 
}

@article{Crouse2016,
  title = {On implementing 2D rectangular assignment algorithms},
  volume = {52},
  ISSN = {0018-9251},
  url = {http://dx.doi.org/10.1109/TAES.2016.140952},
  DOI = {10.1109/taes.2016.140952},
  number = {4},
  journal = {IEEE Trans. Aerosp. Electron. Syst.},
  publisher = {Institute of Electrical and Electronics Engineers (IEEE)},
  author = {Crouse,  David F.},
  year = {2016},
  month = aug,
  pages = {1679–1696}
}

@article{Vlachas2021,
  title = {Accelerated Simulations of Molecular Systems through Learning of Effective Dynamics},
  volume = {18},
  ISSN = {1549-9626},
  url = {http://dx.doi.org/10.1021/acs.jctc.1c00809},
  DOI = {10.1021/acs.jctc.1c00809},
  number = {1},
  journal = {J. Chem. Theory Comput.},
  publisher = {American Chemical Society (ACS)},
  author = {Vlachas,  Pantelis R. and Zavadlav,  Julija and Praprotnik,  Matej and Koumoutsakos,  Petros},
  year = {2021},
  month = dec,
  pages = {538–549}
}

@inproceedings{bowenjing_torsional,
 author = {Jing, Bowen and Corso, Gabriele and Chang, Jeffrey and Barzilay, Regina and Jaakkola, Tommi},
 booktitle = {Advances in Neural Information Processing Systems},
 editor = {S. Koyejo and S. Mohamed and A. Agarwal and D. Belgrave and K. Cho and A. Oh},
 pages = {24240--24253},
 publisher = {Curran Associates, Inc.},
 title = {Torsional Diffusion for Molecular Conformer Generation},
 url = {https://proceedings.neurips.cc/paper_files/paper/2022/file/994545b2308bbbbc97e3e687ea9e464f-Paper-Conference.pdf},
 volume = {35},
 year = {2022}
}

@article{Wang2004,
  title = {Development and testing of a general amber force field},
  volume = {25},
  ISSN = {1096-987X},
  url = {http://dx.doi.org/10.1002/jcc.20035},
  DOI = {10.1002/jcc.20035},
  number = {9},
  journal = {J. Comput. Chem.},
  publisher = {Wiley},
  author = {Wang,  Junmei and Wolf,  Romain M. and Caldwell,  James W. and Kollman,  Peter A. and Case,  David A.},
  year = {2004},
  month = apr,
  pages = {1157–1174}
}

@ARTICLE{rupp,
  author  = {M. Rupp and A. Tkatchenko and K.-R. M\"uller and O. A. von Lilienfeld},
  title   = {Fast and accurate modeling of molecular atomization energies with machine learning},
  journal = "Phys. Rev. Lett.",
  year    = 2012,
  volume  = 108,
  pages   = 058301,
  doi ={10.1103/PhysRevLett.108.058301}
}

@article{Kabsch1976,
  title = {A solution for the best rotation to relate two sets of vectors},
  volume = {32},
  ISSN = {0567-7394},
  url = {http://dx.doi.org/10.1107/S0567739476001873},
  DOI = {10.1107/s0567739476001873},
  number = {5},
  journal = {Acta Crystallogr. A},
  publisher = {International Union of Crystallography (IUCr)},
  author = {Kabsch,  W.},
  year = {1976},
  month = sep,
  pages = {922–923}
}

@book{Leimkuhler2015,
  title = {Molecular Dynamics: With Deterministic and Stochastic Numerical Methods},
  ISBN = {9783319163758},
  ISSN = {2196-9973},
  url = {http://dx.doi.org/10.1007/978-3-319-16375-8},
  DOI = {10.1007/978-3-319-16375-8},
  journal = {Interdisciplinary Applied Mathematics},
  publisher = {Springer International Publishing},
  author = {Leimkuhler,  Ben and Matthews,  Charles},
  year = {2015}
}

@article{Karplus2002,
  title = {Molecular dynamics simulations of biomolecules},
  volume = {9},
  ISSN = {1072-8368},
  url = {http://dx.doi.org/10.1038/nsb0902-646},
  DOI = {10.1038/nsb0902-646},
  number = {9},
  journal = {NNat. Struct. Biol.},
  publisher = {Springer Science and Business Media LLC},
  author = {Karplus,  Martin and McCammon,  J. Andrew},
  year = {2002},
  month = sep,
  pages = {646–652}
}

@article{Kabylda2025,
  title = {Molecular Simulations with a Pretrained Neural Network and Universal Pairwise Force Fields},
  ISSN = {1520-5126},
  url = {http://dx.doi.org/10.1021/jacs.5c09558},
  DOI = {10.1021/jacs.5c09558},
  journal = {J. Am. Chem. Soc.},
  publisher = {American Chemical Society (ACS)},
  author = {Kabylda,  Adil and Frank,  J. Thorben and Suárez-Dou,  Sergio and Khabibrakhmanov,  Almaz and Medrano Sandonas,  Leonardo and Unke,  Oliver T. and Chmiela,  Stefan and M\"{u}ller,  Klaus-Robert and Tkatchenko,  Alexandre},
  year = {2025},
  month = aug 
}

@article{Voter1997,
  title = {Hyperdynamics: Accelerated Molecular Dynamics of Infrequent Events},
  volume = {78},
  ISSN = {1079-7114},
  url = {http://dx.doi.org/10.1103/PhysRevLett.78.3908},
  DOI = {10.1103/physrevlett.78.3908},
  number = {20},
  journal = {Phys. Rev. Lett.},
  publisher = {American Physical Society (APS)},
  author = {Voter,  Arthur F.},
  year = {1997},
  month = may,
  pages = {3908–3911}
}

@article{PalmerIII2001,
  title = {NMR Probes of Molecular Dynamics: Overview and Comparison with Other Techniques},
  volume = {30},
  ISSN = {1545-4266},
  url = {http://dx.doi.org/10.1146/annurev.biophys.30.1.129},
  DOI = {10.1146/annurev.biophys.30.1.129},
  number = {1},
  journal = {Annu. Rev. Biophys. Biomol. Struct.},
  publisher = {Annual Reviews},
  author = {Palmer III,  Arthur G.},
  year = {2001},
  month = jun,
  pages = {129–155}
}

@article{Schuler2013,
  title = {Single-molecule FRET of protein structure and dynamics - a primer},
  volume = {11},
  ISSN = {1477-3155},
  url = {http://dx.doi.org/10.1186/1477-3155-11-S1-S2},
  DOI = {10.1186/1477-3155-11-s1-s2},
  number = {Suppl 1},
  journal = {J. Nanobiotechnol.},
  publisher = {Springer Science and Business Media LLC},
  author = {Schuler,  Benjamin},
  year = {2013},
  pages = {S2}
}

@book{Risken1996,
  title = {The Fokker-Planck Equation: Methods of Solution and Applications},
  ISBN = {9783642615443},
  ISSN = {0172-7389},
  url = {http://dx.doi.org/10.1007/978-3-642-61544-3},
  DOI = {10.1007/978-3-642-61544-3},
  journal = {Springer Series in Synergetics},
  publisher = {Springer Berlin Heidelberg},
  author = {Risken,  Hannes},
  year = {1996}
}

@article{sma-md,
  title = {Generation of conformational ensembles of small molecules via surrogate model-assisted molecular dynamics},
  volume = {5},
  ISSN = {2632-2153},
  url = {http://dx.doi.org/10.1088/2632-2153/ad3b64},
  DOI = {10.1088/2632-2153/ad3b64},
  number = {2},
  journal = { Mach. learn.: sci. technol.},
  publisher = {IOP Publishing},
  author = {Viguera Diez,  Juan and Romeo Atance,  Sara and Engkvist,  Ola and Olsson,  Simon},
  year = {2024},
  month = apr,
  pages = {025010}
}

@inproceedings{transbgs,
 author = {Klein, Leon and No\'{e}, Frank},
 booktitle = {Advances in Neural Information Processing Systems},
 editor = {A. Globerson and L. Mackey and D. Belgrave and A. Fan and U. Paquet and J. Tomczak and C. Zhang},
 pages = {45281--45314},
 publisher = {Curran Associates, Inc.},
 title = {Transferable Boltzmann Generators},
 url = {https://proceedings.neurips.cc/paper_files/paper/2024/file/5035a409f5798e188079e236f437e522-Paper-Conference.pdf},
 volume = {37},
 year = {2024}
}

@inproceedings{ito,
 author = {Schreiner, Mathias and Winther, Ole and Olsson, Simon},
 booktitle = {Advances in Neural Information Processing Systems},
 editor = {A. Oh and T. Naumann and A. Globerson and K. Saenko and M. Hardt and S. Levine},
 pages = {36449--36462},
 publisher = {Curran Associates, Inc.},
 title = {Implicit Transfer Operator Learning: Multiple Time-Resolution Models for Molecular Dynamics},
 url = {https://proceedings.neurips.cc/paper_files/paper/2023/file/7274ed909a312d4d869cc328ad1c5f04-Paper-Conference.pdf},
 volume = {36},
 year = {2023}
}

@inproceedings{timewarp,
 author = {Klein, Leon and Foong, Andrew and Fjelde, Tor and Mlodozeniec, Bruno and Brockschmidt, Marc and Nowozin, Sebastian and Noe, Frank and Tomioka, Ryota},
 booktitle = {Advances in Neural Information Processing Systems},
 editor = {A. Oh and T. Naumann and A. Globerson and K. Saenko and M. Hardt and S. Levine},
 pages = {52863--52883},
 publisher = {Curran Associates, Inc.},
 title = {Timewarp: Transferable Acceleration of Molecular Dynamics by Learning Time-Coarsened Dynamics},
 url = {https://proceedings.neurips.cc/paper_files/paper/2023/file/a598c367280f9054434fdcc227ce4d38-Paper-Conference.pdf},
 volume = {36},
 year = {2023}
}

@article{Wang2024,
  title = {From Latent Dynamics to Meaningful Representations},
  volume = {20},
  ISSN = {1549-9626},
  url = {http://dx.doi.org/10.1021/acs.jctc.4c00249},
  DOI = {10.1021/acs.jctc.4c00249},
  number = {9},
  journal = {J. Chem. Theory Comput.},
  publisher = {American Chemical Society (ACS)},
  author = {Wang,  Dedi and Wang,  Yihang and Evans,  Luke and Tiwary,  Pratyush},
  year = {2024},
  month = apr,
  pages = {3503–3513}
}

@article{schutte2009conformation,
  title={Conformation dynamics},
  author={Sch{\"u}tte, Christof and Noe, Frank and Meerbach, Eike and Metzner, Philipp and Hartmann, Carsten},
  journal={Proc. Int. Congr. ICIAM},
  pages={297--336},
  year={2009}
}

@article{Langevin,
title="Sur la th\'eorie du mouvement brownien", 
author={Langevin, Paul},
journal="C. R. Acad. Sci. (Paris)",
volume="146",
pages={530–-533},
year={1908}}

@inproceedings{1806.07366,
 author = {Chen, Ricky T. Q. and Rubanova, Yulia and Bettencourt, Jesse and Duvenaud, David K},
 booktitle = {Advances in Neural Information Processing Systems},
 editor = {S. Bengio and H. Wallach and H. Larochelle and K. Grauman and N. Cesa-Bianchi and R. Garnett},
 pages = {},
 publisher = {Curran Associates, Inc.},
 title = {Neural Ordinary Differential Equations},
 url = {https://proceedings.neurips.cc/paper_files/paper/2018/file/69386f6bb1dfed68692a24c8686939b9-Paper.pdf},
 volume = {31},
 year = {2018}
}

@article{Ross2020,
  title = {Enhancing Water Sampling in Free Energy Calculations with Grand Canonical Monte Carlo},
  volume = {16},
  ISSN = {1549-9626},
  url = {http://dx.doi.org/10.1021/acs.jctc.0c00660},
  DOI = {10.1021/acs.jctc.0c00660},
  number = {10},
  journal = {J. Chem. Theory Comput.},
  publisher = {American Chemical Society (ACS)},
  author = {Ross,  Gregory A. and Russell,  Ellery and Deng,  Yuqing and Lu,  Chao and Harder,  Edward D. and Abel,  Robert and Wang,  Lingle},
  year = {2020},
  month = sep,
  pages = {6061–6076}
}

@article{Nilmeier2011,
  title = {Nonequilibrium candidate Monte Carlo is an efficient tool for equilibrium simulation},
  volume = {108},
  ISSN = {1091-6490},
  url = {http://dx.doi.org/10.1073/pnas.1106094108},
  DOI = {10.1073/pnas.1106094108},
  number = {45},
  journal = {Proc. Natl. Acad. Sci. U.S.A.},
  publisher = {Proceedings of the National Academy of Sciences},
  author = {Nilmeier,  Jerome P. and Crooks,  Gavin E. and Minh,  David D. L. and Chodera,  John D.},
  year = {2011},
  month = oct 
}

@article{Nske2014,
  title = {Variational Approach to Molecular Kinetics},
  volume = {10},
  ISSN = {1549-9626},
  url = {http://dx.doi.org/10.1021/ct4009156},
  DOI = {10.1021/ct4009156},
  number = {4},
  journal = {J. Chem. Theory Comput.},
  publisher = {American Chemical Society (ACS)},
  author = {N\"{u}ske,  Feliks and Keller,  Bettina G. and Pérez-Hernández,  Guillermo and Mey,  Antonia S. J. S. and Noé,  Frank},
  year = {2014},
  month = mar,
  pages = {1739–1752}
}

@article{Lin1991,
  title = {Divergence measures based on the Shannon entropy},
  volume = {37},
  ISSN = {0018-9448},
  url = {http://dx.doi.org/10.1109/18.61115},
  DOI = {10.1109/18.61115},
  number = {1},
  journal = {IEEE Transactions on Information Theory},
  publisher = {Institute of Electrical and Electronics Engineers (IEEE)},
  author = {Lin,  J.},
  year = {1991},
  pages = {145–151}
}

@article{Sidky2020,
  title = {Molecular latent space simulators},
  volume = {11},
  ISSN = {2041-6539},
  url = {http://dx.doi.org/10.1039/D0SC03635H},
  DOI = {10.1039/d0sc03635h},
  number = {35},
  journal = {Chem. Sci.},
  publisher = {Royal Society of Chemistry (RSC)},
  author = {Sidky,  Hythem and Chen,  Wei and Ferguson,  Andrew L.},
  year = {2020},
  pages = {9459–9467}
}

@article{Prinz2011,
  doi = {10.1063/1.3565032},
  url = {https://doi.org/10.1063/1.3565032},
  year = {2011},
  month = may,
  publisher = {{AIP} Publishing},
  volume = {134},
  number = {17},
  pages = {174105},
  author = {Jan-Hendrik Prinz and Hao Wu and Marco Sarich and Bettina Keller and Martin Senne and Martin Held and John D. Chodera and Christof Sch\"{u}tte and Frank No{\'{e}}},
  title = {Markov models of molecular kinetics: Generation and validation},
  journal = {J. Chem. Phys.}
}

@article{Swope2004,
  doi = {10.1021/jp037421y},
  url = {https://doi.org/10.1021/jp037421y},
  year = {2004},
  month = apr,
  publisher = {American Chemical Society ({ACS})},
  volume = {108},
  number = {21},
  pages = {6571--6581},
  author = {William C. Swope and Jed W. Pitera and Frank Suits},
  title = {Describing Protein Folding Kinetics by Molecular Dynamics Simulations. 1. Theory},
  journal = {J. Phys. Chem. B}
}

@article{JMLR:v26:23-1605,
  author  = {Michael Albergo and Nicholas M. Boffi and Eric Vanden-Eijnden},
  title   = {Stochastic Interpolants: A Unifying Framework for Flows and Diffusions},
  journal = {J. Mach. Learn. Res.},
  year    = {2025},
  volume  = {26},
  number  = {209},
  pages   = {1--80},
  url     = {http://jmlr.org/papers/v26/23-1605.html}
}

@article{Ohmura2014,
  title = {MDGRAPE-4: a special-purpose computer system for molecular dynamics simulations},
  volume = {372},
  ISSN = {1471-2962},
  url = {http://dx.doi.org/10.1098/rsta.2013.0387},
  DOI = {10.1098/rsta.2013.0387},
  number = {2021},
  journal = {Phil. Trans. R. Soc. A,},
  publisher = {The Royal Society},
  author = {Ohmura,  Itta and Morimoto,  Gentaro and Ohno,  Yousuke and Hasegawa,  Aki and Taiji,  Makoto},
  year = {2014},
  month = aug,
  pages = {20130387}
}

@article{Tiwary2016,
  title = {Spectral gap optimization of order parameters for sampling complex molecular systems},
  volume = {113},
  ISSN = {1091-6490},
  url = {http://dx.doi.org/10.1073/pnas.1600917113},
  DOI = {10.1073/pnas.1600917113},
  number = {11},
  journal = {Proc. Natl. Acad. Sci. U.S.A.},
  publisher = {Proceedings of the National Academy of Sciences},
  author = {Tiwary,  Pratyush and Berne,  B. J.},
  year = {2016},
  month = feb,
  pages = {2839–2844}
}

@article{Chen2018,
  title = {Molecular enhanced sampling with autoencoders: On‐the‐fly collective variable discovery and accelerated free energy landscape exploration},
  volume = {39},
  ISSN = {1096-987X},
  url = {http://dx.doi.org/10.1002/jcc.25520},
  DOI = {10.1002/jcc.25520},
  number = {25},
  journal = {J. Comput. Chem.},
  publisher = {Wiley},
  author = {Chen,  Wei and Ferguson,  Andrew L.},
  year = {2018},
  month = sep,
  pages = {2079–2102}
}

@article{Shaw2010,
  title = {Atomic-Level Characterization of the Structural Dynamics of Proteins},
  volume = {330},
  ISSN = {1095-9203},
  url = {http://dx.doi.org/10.1126/science.1187409},
  DOI = {10.1126/science.1187409},
  number = {6002},
  journal = {Science},
  publisher = {American Association for the Advancement of Science (AAAS)},
  author = {Shaw,  David E. and Maragakis,  Paul and Lindorff-Larsen,  Kresten and Piana,  Stefano and Dror,  Ron O. and Eastwood,  Michael P. and Bank,  Joseph A. and Jumper,  John M. and Salmon,  John K. and Shan,  Yibing and Wriggers,  Willy},
  year = {2010},
  month = oct,
  pages = {341–346}
}

@book{msmbook,
  author = {Bowman, Gregory R. and Pande, Vijay S. and Noé, Frank},
  title = {An Introduction to Markov State Models and Their Application to Long Timescale Molecular Simulation},
  publisher = {Springer},
  address = {Dordrecht, Netherlands},
  year = {2014},
  doi = {10.1007/978-94-007-7606-7},
  isbn = {978-94-007-7606-7}
}

@article{Case2023,
  title = {AmberTools},
  volume = {63},
  ISSN = {1549-960X},
  url = {http://dx.doi.org/10.1021/acs.jcim.3c01153},
  DOI = {10.1021/acs.jcim.3c01153},
  number = {20},
  journal = {J. Chem. Inf. Model.},
  publisher = {American Chemical Society (ACS)},
  author = {Case,  David A. and Aktulga,  Hasan Metin and Belfon,  Kellon and Cerutti,  David S. and Cisneros,  G. Andrés and Cruzeiro,  Vinícius Wilian D. and Forouzesh,  Negin and Giese,  Timothy J. and G\"{o}tz,  Andreas W. and Gohlke,  Holger and Izadi,  Saeed and Kasavajhala,  Koushik and Kaymak,  Mehmet C. and King,  Edward and Kurtzman,  Tom and Lee,  Tai-Sung and Li,  Pengfei and Liu,  Jian and Luchko,  Tyler and Luo,  Ray and Manathunga,  Madushanka and Machado,  Matias R. and Nguyen,  Hai Minh and O’Hearn,  Kurt A. and Onufriev,  Alexey V. and Pan,  Feng and Pantano,  Sergio and Qi,  Ruxi and Rahnamoun,  Ali and Risheh,  Ali and Schott-Verdugo,  Stephan and Shajan,  Akhil and Swails,  Jason and Wang,  Junmei and Wei,  Haixin and Wu,  Xiongwu and Wu,  Yongxian and Zhang,  Shi and Zhao,  Shiji and Zhu,  Qiang and Cheatham,  Thomas E. and Roe,  Daniel R. and Roitberg,  Adrian and Simmerling,  Carlos and York,  Darrin M. and Nagan,  Maria C. and Merz,  Kenneth M.},
  year = {2023},
  month = oct,
  pages = {6183–6191}
}

@article{Harvey2009,
  title = {ACEMD: Accelerating Biomolecular Dynamics in the Microsecond Time Scale},
  volume = {5},
  ISSN = {1549-9626},
  url = {http://dx.doi.org/10.1021/ct9000685},
  DOI = {10.1021/ct9000685},
  number = {6},
  journal = {J. Chem. Theory Comput.},
  publisher = {American Chemical Society (ACS)},
  author = {Harvey,  M. J. and Giupponi,  G. and Fabritiis,  G. De},
  year = {2009},
  month = may,
  pages = {1632–1639}
}

@article{Abraham2015,
  title = {GROMACS: High performance molecular simulations through multi-level parallelism from laptops to supercomputers},
  volume = {1–2},
  ISSN = {2352-7110},
  url = {http://dx.doi.org/10.1016/j.softx.2015.06.001},
  DOI = {10.1016/j.softx.2015.06.001},
  journal = {SoftwareX},
  publisher = {Elsevier BV},
  author = {Abraham,  Mark James and Murtola,  Teemu and Schulz,  Roland and Páll,  Szilárd and Smith,  Jeremy C. and Hess,  Berk and Lindahl,  Erik},
  year = {2015},
  month = sep,
  pages = {19–25}
}

@article{Eastman2023,
  title = {OpenMM 8: Molecular Dynamics Simulation with Machine Learning Potentials},
  volume = {128},
  ISSN = {1520-5207},
  url = {http://dx.doi.org/10.1021/acs.jpcb.3c06662},
  DOI = {10.1021/acs.jpcb.3c06662},
  number = {1},
  journal = {J. Phys. Chem. B},
  publisher = {American Chemical Society (ACS)},
  author = {Eastman,  Peter and Galvelis,  Raimondas and Peláez,  Raúl P. and Abreu,  Charlles R. A. and Farr,  Stephen E. and Gallicchio,  Emilio and Gorenko,  Anton and Henry,  Michael M. and Hu,  Frank and Huang,  Jing and Kr\"{a}mer,  Andreas and Michel,  Julien and Mitchell,  Joshua A. and Pande,  Vijay S. and Rodrigues,  João PGLM and Rodriguez-Guerra,  Jaime and Simmonett,  Andrew C. and Singh,  Sukrit and Swails,  Jason and Turner,  Philip and Wang,  Yuanqing and Zhang,  Ivy and Chodera,  John D. and De Fabritiis,  Gianni and Markland,  Thomas E.},
  year = {2023},
  month = dec,
  pages = {109–116}
}

@article{Shirts2000,
  title = {Screen Savers of the World Unite!},
  volume = {290},
  ISSN = {1095-9203},
  url = {http://dx.doi.org/10.1126/science.290.5498.1903},
  DOI = {10.1126/science.290.5498.1903},
  number = {5498},
  journal = {Science},
  publisher = {American Association for the Advancement of Science (AAAS)},
  author = {Shirts,  Michael and Pande,  Vijay S.},
  year = {2000},
  month = dec,
  pages = {1903–1904}
}

@inproceedings{Shaw2021,
  series = {SC ’21},
  title = {Anton 3: twenty microseconds of molecular dynamics simulation before lunch},
  url = {http://dx.doi.org/10.1145/3458817.3487397},
  DOI = {10.1145/3458817.3487397},
  booktitle = {Proceedings of the International Conference for High Performance Computing,  Networking,  Storage and Analysis},
  publisher = {ACM},
  author = {Shaw,  David E. and Adams,  Peter J. and Azaria,  Asaph and Bank,  Joseph A. and Batson,  Brannon and Bell,  Alistair and Bergdorf,  Michael and Bhatt,  Jhanvi and Butts,  J. Adam and Correia,  Timothy and Dirks,  Robert M. and Dror,  Ron O. and Eastwood,  Michael P. and Edwards,  Bruce and Even,  Amos and Feldmann,  Peter and Fenn,  Michael and Fenton,  Christopher H. and Forte,  Anthony and Gagliardo,  Joseph and Gill,  Gennette and Gorlatova,  Maria and Greskamp,  Brian and Grossman,  J.P. and Gullingsrud,  Justin and Harper,  Anissa and Hasenplaugh,  William and Heily,  Mark and Heshmat,  Benjamin Colin and Hunt,  Jeremy and Ierardi,  Douglas J. and Iserovich,  Lev and Jackson,  Bryan L. and Johnson,  Nick P. and Kirk,  Mollie M. and Klepeis,  John L. and Kuskin,  Jeffrey S. and Mackenzie,  Kenneth M. and Mader,  Roy J. and McGowen,  Richard and McLaughlin,  Adam and Moraes,  Mark A. and Nasr,  Mohamed H. and Nociolo,  Lawrence J. and O’Donnell,  Lief and Parker,  Andrew and Peticolas,  Jon L. and Pocina,  Goran and Predescu,  Cristian and Quan,  Terry and Salmon,  John K. and Schwink,  Carl and Shim,  Keun Sup and Siddique,  Naseer and Spengler,  Jochen and Szalay,  Tamas and Tabladillo,  Raymond and Tartler,  Reinhard and Taube,  Andrew G. and Theobald,  Michael and Towles,  Brian and Vick,  William and Wang,  Stanley C. and Wazlowski,  Michael and Weingarten,  Madeleine J. and Williams,  John M. and Yuh,  Kevin A.},
  year = {2021},
  month = nov,
  pages = {1–11},
  collection = {SC ’21}
}

@article{Henin2022,
Author = {Jérôme Hénin and Tony Lelièvre and Michael R. Shirts and Omar Valsson and Lucie Delemotte},
Title = {Enhanced sampling methods for molecular dynamics simulations},
Year = {2022},
journal = {Living J. Comput. Mol. Sci},
volume = 4,
number = 1,
page = 1583,
Eprint = {arXiv:2202.04164},
Howpublished = {Living J. Comput. Mol. Sci, 4(1), 1583 (2022)},
Doi = {10.33011/livecoms.4.1.1583},
}

@article{Buchete2008,
  title = {Coarse Master Equations for Peptide Folding Dynamics},
  volume = {112},
  ISSN = {1520-5207},
  url = {http://dx.doi.org/10.1021/jp0761665},
  DOI = {10.1021/jp0761665},
  number = {19},
  journal = {J. Phys. Chem. B},
  publisher = {American Chemical Society (ACS)},
  author = {Buchete,  Nicolae-Viorel and Hummer,  Gerhard},
  year = {2008},
  month = jan,
  pages = {6057–6069}
}

@misc{2502.13797,
Author = {Alessandro Caruso and Jacopo Venturin and Lorenzo Giambagli and Edoardo Rolando and Frank Noé and Cecilia Clementi},
Title = {Extending the RANGE of Graph Neural Networks: Relaying Attention Nodes for Global Encoding},
Year = {2025},
Eprint = {arXiv:2502.13797},
url={https://doi.org/10.48550/arXiv.2502.13797}
}

@inproceedings{NEURIPS2024_fad8e191,
 author = {Qu, Eric and Krishnapriyan, Aditi S.},
 booktitle = {Advances in Neural Information Processing Systems},
 editor = {A. Globerson and L. Mackey and D. Belgrave and A. Fan and U. Paquet and J. Tomczak and C. Zhang},
 pages = {139030--139053},
 publisher = {Curran Associates, Inc.},
 title = {The Importance of Being Scalable: Improving the Speed and Accuracy of Neural Network Interatomic Potentials Across Chemical Domains},
 url = {https://proceedings.neurips.cc/paper_files/paper/2024/file/fad8e1915f66161581bb127ccf01092e-Paper-Conference.pdf},
 volume = {37},
 year = {2024}
}

@misc{2503.23794,
Author = {Fabian L. Thiemann and Thiago Reschützegger and Massimiliano Esposito and Tseden Taddese and Juan D. Olarte-Plata and Fausto Martelli},
Title = {Force-Free Molecular Dynamics Through Autoregressive Equivariant Networks},
Year = {2025},
Book = {arXiv:2503.23794},
URL = {https://doi.org/10.48550/arXiv.2503.23794}
}

@article{Feenstra1999,
  title = {Improving efficiency of large time-scale molecular dynamics simulations of hydrogen-rich systems},
  volume = {20},
  ISSN = {1096-987X},
  url = {http://dx.doi.org/10.1002/(SICI)1096-987X(199906)20:8<786::AID-JCC5>3.0.CO;2-B},
  DOI = {10.1002/(sici)1096-987x(199906)20:8<786::aid-jcc5>3.0.co;2-b},
  number = {8},
  journal = {J. Comput. Chem.},
  publisher = {Wiley},
  author = {Feenstra,  K. Anton and Hess,  Berk and Berendsen,  Herman J. C.},
  year = {1999},
  month = jun,
  pages = {786–798}
}

@article{Leimkuhler2013,
  title = {Robust and efficient configurational molecular sampling via Langevin dynamics},
  volume = {138},
  ISSN = {1089-7690},
  url = {http://dx.doi.org/10.1063/1.4802990},
  DOI = {10.1063/1.4802990},
  number = {17},
  journal = {J. Chem. Phys.},
  publisher = {AIP Publishing},
  author = {Leimkuhler,  Benedict and Matthews,  Charles},
  year = {2013},
  month = may 
}

@article{Tuckerman1992,
  title = {Reversible multiple time scale molecular dynamics},
  volume = {97},
  ISSN = {1089-7690},
  url = {http://dx.doi.org/10.1063/1.463137},
  DOI = {10.1063/1.463137},
  number = {3},
  journal = {J. Chem. Phys.},
  publisher = {AIP Publishing},
  author = {Tuckerman,  M. and Berne,  B. J. and Martyna,  G. J.},
  year = {1992},
  month = aug,
  pages = {1990–2001}
}

@article{Batatia2025,
  title = {A foundation model for atomistic materials chemistry},
  volume = {163},
  ISSN = {1089-7690},
  url = {http://dx.doi.org/10.1063/5.0297006},
  DOI = {10.1063/5.0297006},
  number = {18},
  journal = {J. Chem. Phys.},
  publisher = {AIP Publishing},
  author = {Batatia,  Ilyes and Benner,  Philipp and Chiang,  Yuan and Elena,  Alin M. and Kovács,  Dávid P. and Riebesell,  Janosh and Advincula,  Xavier R. and Asta,  Mark and Avaylon,  Matthew and Baldwin,  William J. and Berger,  Fabian and Bernstein,  Noam and Bhowmik,  Arghya and Bigi,  Filippo and Blau,  Samuel M. and Cărare,  Vlad and Ceriotti,  Michele and Chong,  Sanggyu and Darby,  James P. and De,  Sandip and Della Pia,  Flaviano and Deringer,  Volker L. and Elijošius,  Rokas and El-Machachi,  Zakariya and Fako,  Edvin and Falcioni,  Fabio and Ferrari,  Andrea C. and Gardner,  John L. A. and Gawkowski,  Mikołaj J. and Genreith-Schriever,  Annalena and George,  Janine and Goodall,  Rhys E. A. and Grandel,  Jonas and Grey,  Clare P. and Grigorev,  Petr and Han,  Shuang and Handley,  Will and Heenen,  Hendrik H. and Hermansson,  Kersti and Ho,  Cheuk Hin and Hofmann,  Stephan and Holm,  Christian and Jaafar,  Jad and Jakob,  Konstantin S. and Jung,  Hyunwook and Kapil,  Venkat and Kaplan,  Aaron D. and Karimitari,  Nima and Kermode,  James R. and Kourtis,  Panagiotis and Kroupa,  Namu and Kullgren,  Jolla and Kuner,  Matthew C. and Kuryla,  Domantas and Liepuoniute,  Guoda and Lin,  Chen and Margraf,  Johannes T. and Magdău,  Ioan-Bogdan and Michaelides,  Angelos and Moore,  J. Harry and Naik,  Aakash A. and Niblett,  Samuel P. and Norwood,  Sam Walton and O’Neill,  Niamh and Ortner,  Christoph and Persson,  Kristin A. and Reuter,  Karsten and Rosen,  Andrew S. and Rosset,  Louise A. M. and Schaaf,  Lars L. and Schran,  Christoph and Shi,  Benjamin X. and Sivonxay,  Eric and Stenczel,  Tamás K. and Sutton,  Christopher and Svahn,  Viktor and Swinburne,  Thomas D. and Tilly,  Jules and van der Oord,  Cas and Vargas,  Santiago and Varga-Umbrich,  Eszter and Vegge,  Tejs and Vondrák,  Martin and Wang,  Yangshuai and Witt,  William C. and Wolf,  Thomas and Zills,  Fabian and Csányi,  Gábor},
  year = {2025},
  month = nov 
}

@article{Majewski2023,
  title = {Machine learning coarse-grained potentials of protein thermodynamics},
  volume = {14},
  ISSN = {2041-1723},
  url = {http://dx.doi.org/10.1038/s41467-023-41343-1},
  DOI = {10.1038/s41467-023-41343-1},
  number = {1},
  journal = {Nat. Commun.},
  publisher = {Springer Science and Business Media LLC},
  author = {Majewski,  Maciej and Pérez,  Adrià and Th\"{o}lke,  Philipp and Doerr,  Stefan and Charron,  Nicholas E. and Giorgino,  Toni and Husic,  Brooke E. and Clementi,  Cecilia and Noé,  Frank and De Fabritiis,  Gianni},
  year = {2023},
  month = sep 
}

@article{Charron2025,
  title = {Navigating protein landscapes with a machine-learned transferable coarse-grained model},
  volume = {17},
  ISSN = {1755-4349},
  url = {http://dx.doi.org/10.1038/s41557-025-01874-0},
  DOI = {10.1038/s41557-025-01874-0},
  number = {8},
  journal = {Nat. Chem.},
  publisher = {Springer Science and Business Media LLC},
  author = {Charron,  Nicholas E. and Bonneau,  Klara and Pasos-Trejo,  Aldo S. and Guljas,  Andrea and Chen,  Yaoyi and Musil,  Félix and Venturin,  Jacopo and Gusew,  Daria and Zaporozhets,  Iryna and Kr\"{a}mer,  Andreas and Templeton,  Clark and Kelkar,  Atharva and Durumeric,  Aleksander E. P. and Olsson,  Simon and Pérez,  Adrià and Majewski,  Maciej and Husic,  Brooke E. and Patel,  Ankit and De Fabritiis,  Gianni and Noé,  Frank and Clementi,  Cecilia},
  year = {2025},
  month = jul,
  pages = {1284–1292}
}

@article{ml-coarse-grain,
  title = {Machine Learning of Coarse-Grained Molecular Dynamics Force Fields},
  volume = {5},
  ISSN = {2374-7951},
  url = {http://dx.doi.org/10.1021/acscentsci.8b00913},
  DOI = {10.1021/acscentsci.8b00913},
  number = {5},
  journal = {ACS Cent. Sci.},
  publisher = {American Chemical Society (ACS)},
  author = {Wang,  Jiang and Olsson,  Simon and Wehmeyer,  Christoph and Pérez,  Adrià and Charron,  Nicholas E. and de Fabritiis,  Gianni and Noé,  Frank and Clementi,  Cecilia},
  year = {2019},
  month = apr,
  pages = {755–767}
}

@article{Jung2023,
  title = {Machine-guided path sampling to discover mechanisms of molecular self-organization},
  volume = {3},
  ISSN = {2662-8457},
  url = {http://dx.doi.org/10.1038/s43588-023-00428-z},
  DOI = {10.1038/s43588-023-00428-z},
  number = {4},
  journal = {Nat. Comput. Sci.},
  publisher = {Springer Science and Business Media LLC},
  author = {Jung,  Hendrik and Covino,  Roberto and Arjun,  A. and Leitold,  Christian and Dellago,  Christoph and Bolhuis,  Peter G. and Hummer,  Gerhard},
  year = {2023},
  month = apr,
  pages = {334–345}
}

@book{flory1953principles,
  title={Principles of Polymer Chemistry},
  author={Flory, P.J.},
  isbn={9780801401343},
  lccn={lc53013473},
  series={Baker lectures 1948},
  url={https://books.google.se/books?id=CQ0EbEkT5R0C},
  year={1953},
  publisher={Cornell University Press}
}

@article{Branduardi2007,
  title = {From A to B in free energy space},
  volume = {126},
  ISSN = {1089-7690},
  url = {http://dx.doi.org/10.1063/1.2432340},
  DOI = {10.1063/1.2432340},
  number = {5},
  journal = {J. Chem. Phys.},
  publisher = {AIP Publishing},
  author = {Branduardi,  Davide and Gervasio,  Francesco Luigi and Parrinello,  Michele},
  year = {2007},
  month = feb 
}

@article{PrezHernndez2013,
  title = {Identification of slow molecular order parameters for Markov model construction},
  volume = {139},
  ISSN = {1089-7690},
  url = {http://dx.doi.org/10.1063/1.4811489},
  DOI = {10.1063/1.4811489},
  number = {1},
  journal = {J. Chem. Phys.},
  publisher = {AIP Publishing},
  author = {Pérez-Hernández,  Guillermo and Paul,  Fabian and Giorgino,  Toni and De Fabritiis,  Gianni and Noé,  Frank},
  year = {2013},
  month = jul 
}

@Article{re,
author ="Earl, David J. and Deem, Michael W.",
title  ="Parallel tempering: Theory{,} applications{,} and new perspectives",
journal  ="Phys. Chem. Chem. Phys.",
year  ="2005",
volume  ="7",
issue  ="23",
pages  ="3910-3916",
publisher  ="The Royal Society of Chemistry",
doi  ="10.1039/B509983H",
url  ="http://dx.doi.org/10.1039/B509983H"}

@article{Grubmuller1995,
  title = {Predicting slow structural transitions in macromolecular systems: Conformational flooding},
  author = {Grubm\"uller, Helmut},
  journal = {Phys. Rev. E},
  volume = {52},
  issue = {3},
  pages = {2893--2906},
  numpages = {0},
  year = {1995},
  month = {9},
  publisher = {American Physical Society},
  doi = {10.1103/PhysRevE.52.2893},
  url = {https://link.aps.org/doi/10.1103/PhysRevE.52.2893}
}

@article{Laio2002,
    title = {{Escaping free-energy minima}},
    year = {2002},
    journal = {Proc. Natl. Acad. Sci. U.S.A.},
    author = {Laio, Alessandro and Parrinello, Michele},
    number = {20},
    pages = {12562--12566},
    volume = {99},
    publisher = {National Acad Sciences}
}

@article{Torrie1977,
  doi = {10.1016/0021-9991(77)90121-8},
  url = {https://doi.org/10.1016/0021-9991(77)90121-8},
  year = {1977},
  month = feb,
  publisher = {Elsevier {BV}},
  volume = {23},
  number = {2},
  pages = {187--199},
  author = {G.M. Torrie and J.P. Valleau},
  title = {Nonphysical sampling distributions in Monte Carlo free-energy estimation: Umbrella sampling},
  journal = {J. Comput. Phys.}
}

@inproceedings{
mdgen,
title={Generative Modeling of Molecular Dynamics Trajectories},
author={Bowen Jing and Hannes Stark and Tommi Jaakkola and Bonnie Berger},
booktitle={The Thirty-eighth Annual Conference on Neural Information Processing Systems},
year={2024},
url={https://openreview.net/forum?id=yRRCH1OsGW}
}

@article{bioemu,
  title = {Scalable emulation of protein equilibrium ensembles with generative deep learning},
  volume = {389},
  ISSN = {1095-9203},
  url = {http://dx.doi.org/10.1126/science.adv9817},
  DOI = {10.1126/science.adv9817},
  number = {6761},
  journal = {Science},
  publisher = {American Association for the Advancement of Science (AAAS)},
  author = {Lewis,  Sarah and Hempel,  Tim and Jiménez-Luna,  José and Gastegger,  Michael and Xie,  Yu and Foong,  Andrew Y. K. and Satorras,  Victor García and Abdin,  Osama and Veeling,  Bastiaan S. and Zaporozhets,  Iryna and Chen,  Yaoyi and Yang,  Soojung and Foster,  Adam E. and Schneuing,  Arne and Nigam,  Jigyasa and Barbero,  Federico and Stimper,  Vincent and Campbell,  Andrew and Yim,  Jason and Lienen,  Marten and Shi,  Yu and Zheng,  Shuxin and Schulz,  Hannes and Munir,  Usman and Sordillo,  Roberto and Tomioka,  Ryota and Clementi,  Cecilia and Noé,  Frank},
  year = {2025},
  month = aug 
}

@misc{semlaflow,
      title={SemlaFlow -- Efficient 3D Molecular Generation with Latent Attention and Equivariant Flow Matching}, 
      author={Ross Irwin and Alessandro Tibo and Jon Paul Janet and Simon Olsson},
      year={2025},
      eprint={2406.07266},
      archivePrefix={arXiv},
      primaryClass={cs.LG},
      url={https://arxiv.org/abs/2406.07266}, 
}

@misc{euclideanfastattention,
      title={Euclidean Fast Attention: Machine Learning Global Atomic Representations at Linear Cost}, 
      author={J. Thorben Frank and Stefan Chmiela and Klaus-Robert Müller and Oliver T. Unke},
      year={2024},
      eprint={2412.08541},
      archivePrefix={arXiv},
      primaryClass={cs.LG},
      url={https://arxiv.org/abs/2412.08541}, 
}

@inproceedings{flowmatching,
title={Flow Matching for Generative Modeling},
author={Yaron Lipman and Ricky T. Q. Chen and Heli Ben-Hamu and Maximilian Nickel and Matthew Le},
booktitle={The Eleventh International Conference on Learning Representations },
year={2023},
url={https://openreview.net/forum?id=PqvMRDCJT9t}
}

@article{VAMP,
  title = {Variational Approach for Learning Markov Processes from Time Series Data},
  volume = {30},
  ISSN = {1432-1467},
  url = {http://dx.doi.org/10.1007/s00332-019-09567-y},
  DOI = {10.1007/s00332-019-09567-y},
  number = {1},
  journal = {J. Nonlinear Sci.},
  publisher = {Springer Science and Business Media LLC},
  author = {Wu,  Hao and Noé,  Frank},
  year = {2019},
  month = aug,
  pages = {23–66}
}

@inproceedings{equivariantflowmatching,
 author = {Klein, Leon and Kr\"{a}mer, Andreas and Noe, Frank},
 booktitle = {Advances in Neural Information Processing Systems},
 editor = {A. Oh and T. Naumann and A. Globerson and K. Saenko and M. Hardt and S. Levine},
 pages = {59886--59910},
 publisher = {Curran Associates, Inc.},
 title = {Equivariant flow matching},
 url = {https://proceedings.neurips.cc/paper_files/paper/2023/file/bc827452450356f9f558f4e4568d553b-Paper-Conference.pdf},
 volume = {36},
 year = {2023}
}

@article{thermodynamicinterpolation,
  title = {Thermodynamic Interpolation: A Generative Approach to Molecular Thermodynamics and Kinetics},
  volume = {21},
  ISSN = {1549-9626},
  url = {http://dx.doi.org/10.1021/acs.jctc.4c01557},
  DOI = {10.1021/acs.jctc.4c01557},
  number = {5},
  journal = {J. Chem. Theory Comput.},
  publisher = {American Chemical Society (ACS)},
  author = {Moqvist,  Selma and Chen,  Weilong and Schreiner,  Mathias and N\"{u}ske,  Feliks and Olsson,  Simon},
  year = {2025},
  month = feb,
  pages = {2535–2545}
}

@article{Herron2024,
  title = {Inferring phase transitions and critical exponents from limited observations with thermodynamic maps},
  volume = {121},
  ISSN = {1091-6490},
  url = {http://dx.doi.org/10.1073/pnas.2321971121},
  DOI = {10.1073/pnas.2321971121},
  number = {52},
  journal = {Proc. Natl. Acad. Sci. U.S.A.},
  publisher = {Proceedings of the National Academy of Sciences},
  author = {Herron,  Lukas and Mondal,  Kinjal and Schneekloth,  John S. and Tiwary,  Pratyush},
  year = {2024},
  month = dec 
}

@inproceedings{bopito,
title={Boltzmann priors for Implicit Transfer Operators},
author={Juan Viguera Diez and Mathias Jacob Schreiner and Ola Engkvist and Simon Olsson},
booktitle={The Thirteenth International Conference on Learning Representations},
year={2025},
url={https://openreview.net/forum?id=pRCOZllZdT}
}

@misc{costa2024equijumpproteindynamicssimulation,
      title={EquiJump: Protein Dynamics Simulation via SO(3)-Equivariant Stochastic Interpolants}, 
      author={Allan dos Santos Costa and Ilan Mitnikov and Franco Pellegrini and Ameya Daigavane and Mario Geiger and Zhonglin Cao and Karsten Kreis and Tess Smidt and Emine Kucukbenli and Joseph Jacobson},
      year={2024},
      eprint={2410.09667},
      archivePrefix={arXiv},
      primaryClass={cs.LG},
      url={https://arxiv.org/abs/2410.09667}, 
}

@misc{costa2025acceleratingproteinmoleculardynamics,
      title={Accelerating Protein Molecular Dynamics Simulation with DeepJump}, 
      author={Allan dos Santos Costa and Manvitha Ponnapati and Dana Rubin and Tess Smidt and Joseph Jacobson},
      year={2025},
      eprint={2509.13294},
      archivePrefix={arXiv},
      primaryClass={q-bio.BM},
      url={https://arxiv.org/abs/2509.13294}, 
}

@inproceedings{fuchs2020se3transformers3drototranslationequivariant,
 author = {Fuchs, Fabian and Worrall, Daniel and Fischer, Volker and Welling, Max},
 booktitle = {Advances in Neural Information Processing Systems},
 editor = {H. Larochelle and M. Ranzato and R. Hadsell and M.F. Balcan and H. Lin},
 pages = {1970--1981},
 publisher = {Curran Associates, Inc.},
 title = {SE(3)-Transformers: 3D Roto-Translation Equivariant Attention Networks},
 url = {https://proceedings.neurips.cc/paper_files/paper/2020/file/15231a7ce4ba789d13b722cc5c955834-Paper.pdf},
 volume = {33},
 year = {2020}
}

@InProceedings{katharopoulos2020transformersrnnsfastautoregressive,
  title = 	 {Transformers are {RNN}s: Fast Autoregressive Transformers with Linear Attention},
  author =       {Katharopoulos, Angelos and Vyas, Apoorv and Pappas, Nikolaos and Fleuret, Fran{\c{c}}ois},
  booktitle = 	 {Proceedings of the 37th International Conference on Machine Learning},
  pages = 	 {5156--5165},
  year = 	 {2020},
  editor = 	 {III, Hal Daumé and Singh, Aarti},
  volume = 	 {119},
  series = 	 {Proceedings of Machine Learning Research},
  month = 	 {13--18 Jul},
  publisher =    {PMLR},
  url = 	 {https://proceedings.mlr.press/v119/katharopoulos20a.html}
}

@inproceedings{
bigi2025flashmdlongstrideuniversalprediction,
title={Flash{MD}: long-stride, universal prediction of molecular dynamics},
author={Filippo Bigi and Sanggyu Chong and Agustinus Kristiadi and Michele Ceriotti},
booktitle={The Thirty-ninth Annual Conference on Neural Information Processing Systems},
year={2025},
url={https://openreview.net/forum?id=ogZu06NgQs}
}

@misc{murtada2025mdllm1largelanguagemodel,
      title={MD-LLM-1: A Large Language Model for Molecular Dynamics}, 
      author={Mhd Hussein Murtada and Z. Faidon Brotzakis and Michele Vendruscolo},
      year={2025},
      eprint={2508.03709},
      archivePrefix={arXiv},
      primaryClass={q-bio.BM},
      url={https://arxiv.org/abs/2508.03709}, 
}

@article{Olsson2026,
  title = {Generative molecular dynamics},
  volume = {96},
  ISSN = {0959-440X},
  url = {http://dx.doi.org/10.1016/j.sbi.2025.103213},
  DOI = {10.1016/j.sbi.2025.103213},
  journal = { Curr. Opin. Struct. Biol.},
  publisher = {Elsevier BV},
  author = {Olsson,  Simon},
  year = {2026},
  month = feb,
  pages = {103213}
}

\bibliographystyle{sciencemag}

\newpage

\section*{Supplementary material}
{\Large Transferable Generative Models Bridge Femtosecond to Nanosecond Time-Step Molecular Dynamics}\\
{Juan Viguera Diez, Mathias Schreiner, and Simon Olsson}\\
{\bf Corresponding author:} Simon Olsson (simonols@chalmers.se)\\
\pagebreak
\renewcommand{\thefigure}{S\arabic{figure}}
\renewcommand{\thetable}{S\arabic{table}}
\renewcommand{\figurename}{Supplementary Figure}
\renewcommand{\tablename}{Supplementary Table}
\newcommand{\namedsubsection}[1]{%
    \subsection*{#1}%
    \addcontentsline{toc}{section}{#1}
}
\setcounter{page}{1}

\setcounter{figure}{0}
\setcounter{table}{0}

\etocsetstyle{subsection}
    {}                                          
    {}                                          
    {\noindent\etocname\nobreak\hfill\etocpage\par} 
    {}                                          

\etocsetnexttocdepth{subsection}

\tableofcontents

\setcounter{tocdepth}{3} 
\etocsettocdepth{section}

\pagebreak

\begin{figure}[h!]
    \centering
    \makebox[\textwidth][c]{\includegraphics[width=1\linewidth]{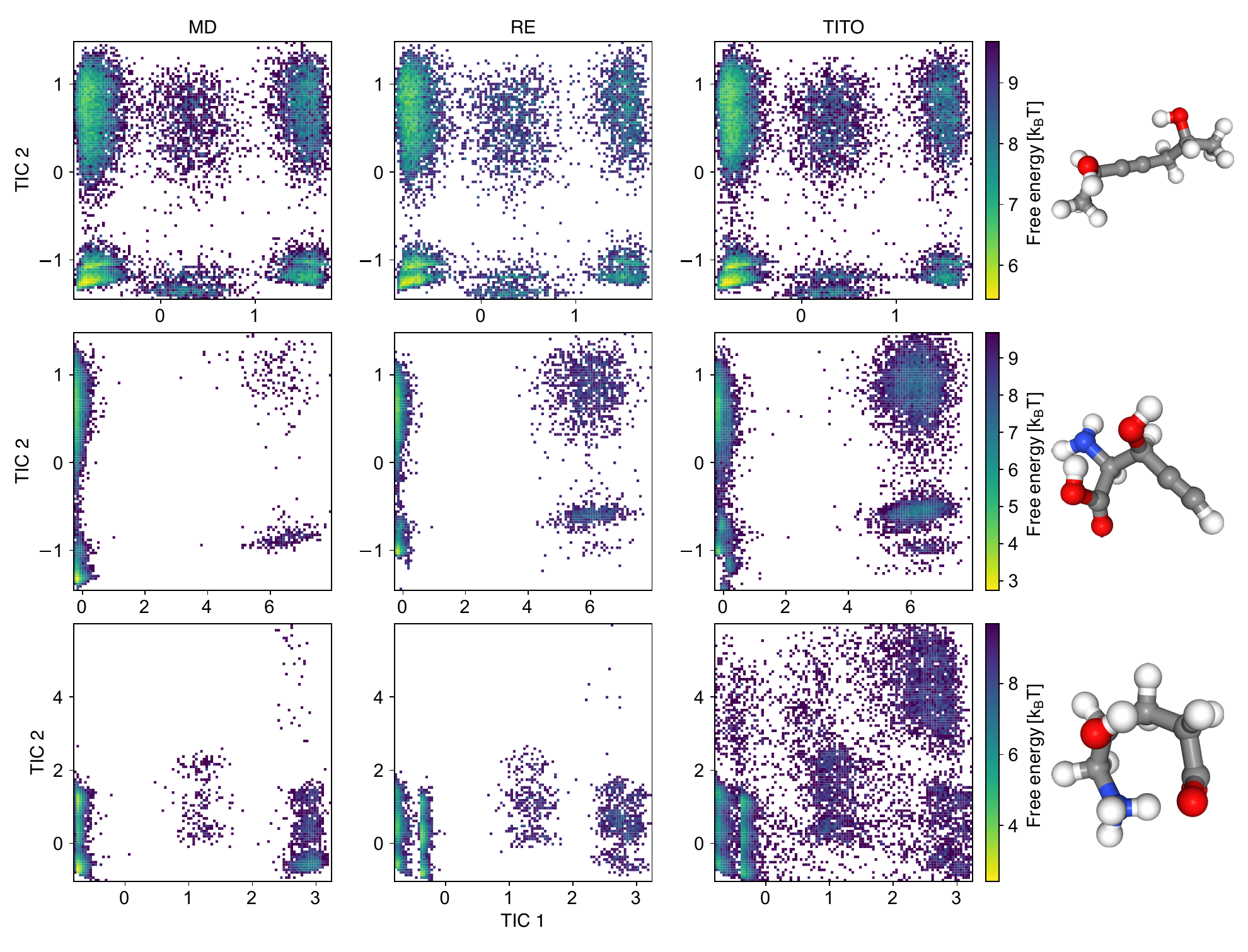}}
    \caption{Free energy landscape comparison between MD, RE and TITO for different regions of JSD. From top to bottom, the JSD are 0.09, 0.21 and 0.35.}
    \label{fig:si-examples-jsds}
\end{figure}

\namedsubsection{Examples of molecules with different Jensen-Shannon divergences}
In Suppl.~Fig. \ref{fig:si-examples-jsds} we compare free energy landscape of MD, RE and TITO for different regions of JSD.

\begin{figure}[h!]
    \centering
    \makebox[\textwidth][c]{\includegraphics[width=1\linewidth]{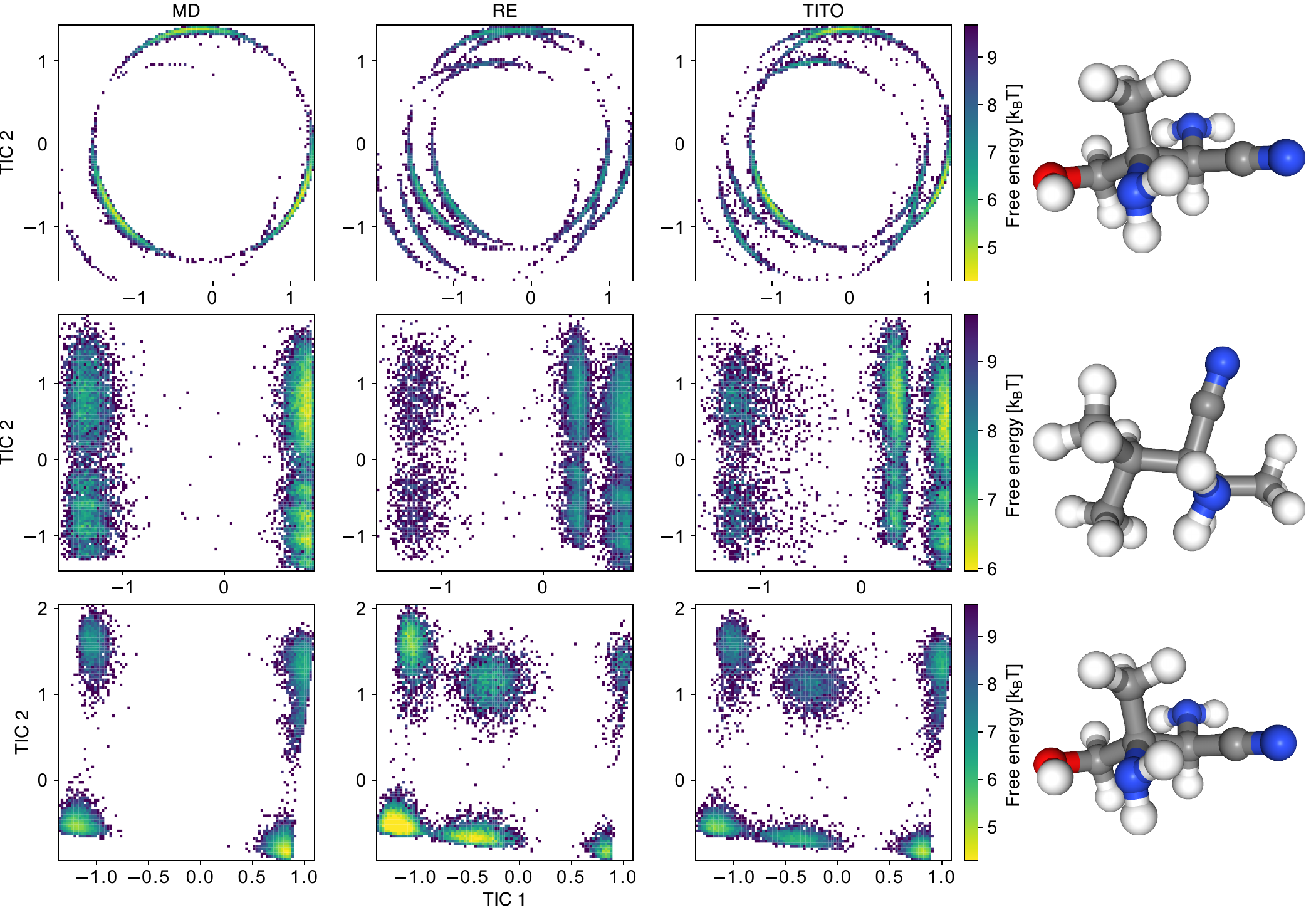}}
    \caption{Free energy landscape comparison between MD, RE and TITO of test set molecules for which TITO recovers states sampled by RE, but not accessible by training set-like MD simulations .}
    \label{fig:si-tito-better-md}
\end{figure}

\namedsubsection{TITO recovers states not accessible by training set-like simulations}
In Suppl.~Fig. \ref{fig:si-tito-better-md} we provide several examples of test molecules for which TITO is able to recover states sampled by RE, which are not accessible by training set-like simulations.

\begin{figure}[h!]
    \centering
    \includegraphics[width=\linewidth]{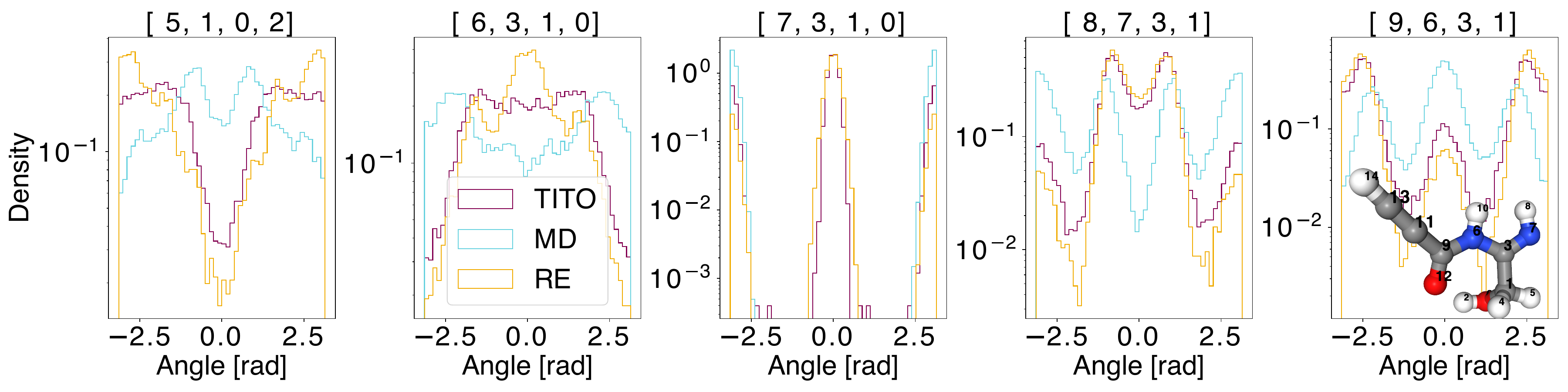}
    \caption{Dihedral angles involved in slowest process of example molecule in Fig.~\ref{fig:long-term} C. MD fails to sample the transition. TITO recovers the equilibrium distribution predicted by RE.}
    \label{fig:si-mol21-torsions}
\end{figure}

\namedsubsection{Overcoming MD timescales with TITO: Structural insights for propiolamide}
\label{sec:si-structural-insights}
In Suppl.~Fig. \ref{fig:si-mol21-torsions} we show that the slowest process of propiolamide (Fig. \ref{fig:long-term}C) involves a dihedral angle sign inversion over the bond 3-1 and a global re-arrangement of other dihedral angles in the molecule. Short MD fails to sample the transition, but TITO samples it and recovers an equilibrium distribution in high agreement with RE.

\begin{figure}[h!]
    \centering
    \includegraphics[width=\linewidth]{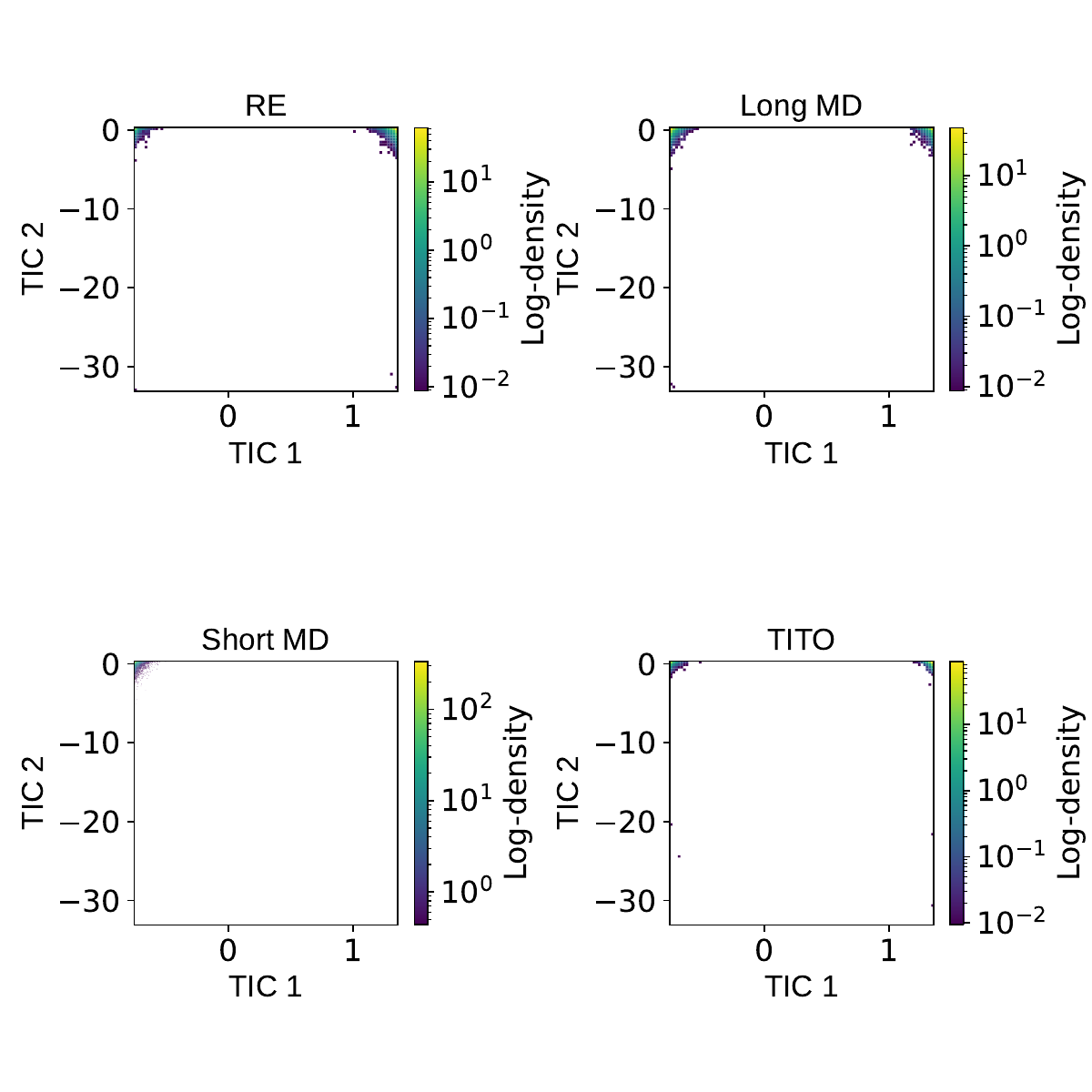}
    \caption{Alternative TICA projections using long MD simulations to estimate TICA models in Fig.~\ref{fig:long-term}C. MD fails to sample the transition. TITO recovers the equilibrium distribution predicted by RE and long MD.}
    \label{fig:si-long-tica}
\end{figure}

\namedsubsection{Alternative TICA projections}
\label{sec:si-alternative-tica}
In Suppl.~Fig. \ref{fig:si-long-tica} we show alternative TICA projections using ultra long MD simulations to estimate TICA models of propiolamide. 

\namedsubsection{Training on different molecules enables state discovery}

\begin{figure}[h!]
    \centering
    \includegraphics[width=1.0\linewidth]{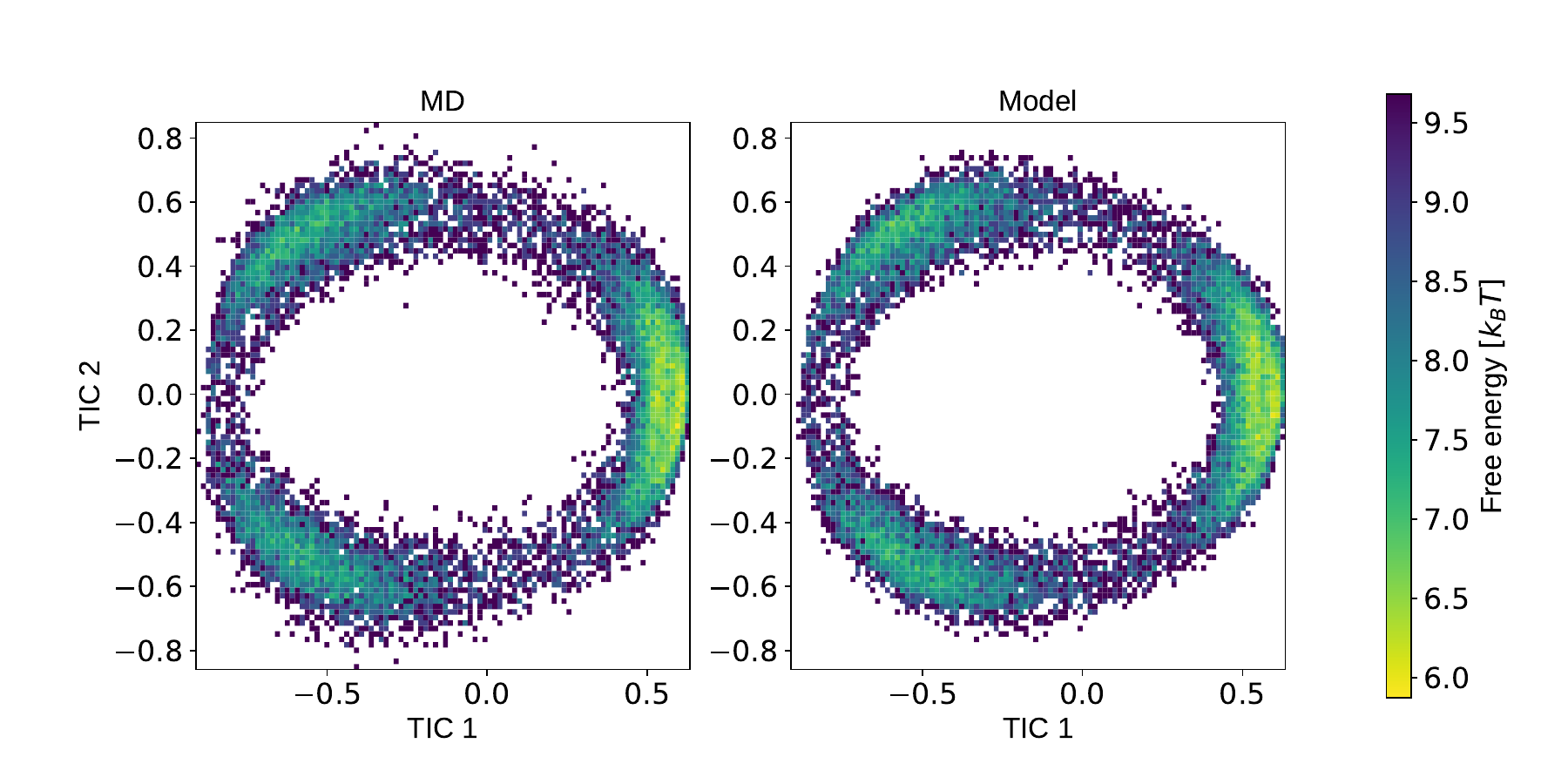}
    \caption{Free energy landscape generated by a model trained only on a training-set-like simulation of the test set propiolamide. The state discovery example presented in Fig. \ref{fig:long-term} C is not recovered. Inference parameters are identical to Fig. \ref{fig:long-term} (TITO).}
    \label{fig:si-one-mol-state-discovery}
\end{figure}

In Suppl.~Fig. \ref{fig:si-one-mol-state-discovery} we show that the state discovery example presented in Fig. \ref{fig:long-term} C is not possible when we train an identical model solely on training-set-like trajectory data from that molecule.

\subsection*{MD fine-tuning time vs JS divergence}
\begin{figure}[h!]
    \centering
    \makebox[\textwidth][c]{\includegraphics[width=0.5\linewidth]{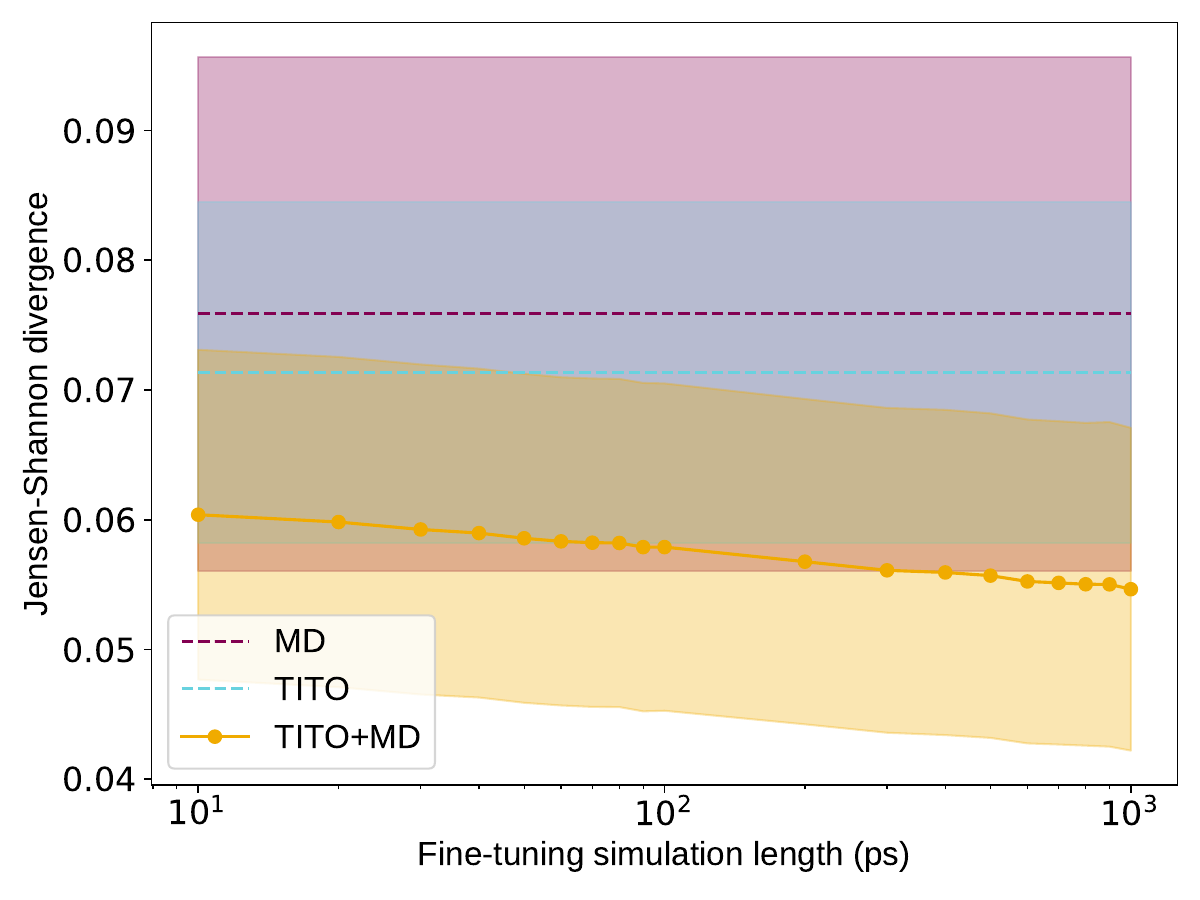}}
    \caption{Mean JS divergence and 95 \% confindence interval for TITO, TITO + short MD and MD versus simulation time applied to sampled from TITO.}
    \label{fig:si-js-vs-mdft}
\end{figure}

In Suppl.~Fig. \ref{fig:si-js-vs-mdft} we show the evolution of average Jensen-Shannon divergence w.r.t. the time that TITO samples are simulated with standard MD. Most of the reduction is achieved during the first 10 $\mathrm{ps}$. 

\begin{figure}[h!]
    \centering
    \makebox[\textwidth][c]{\includegraphics[width=1\linewidth]{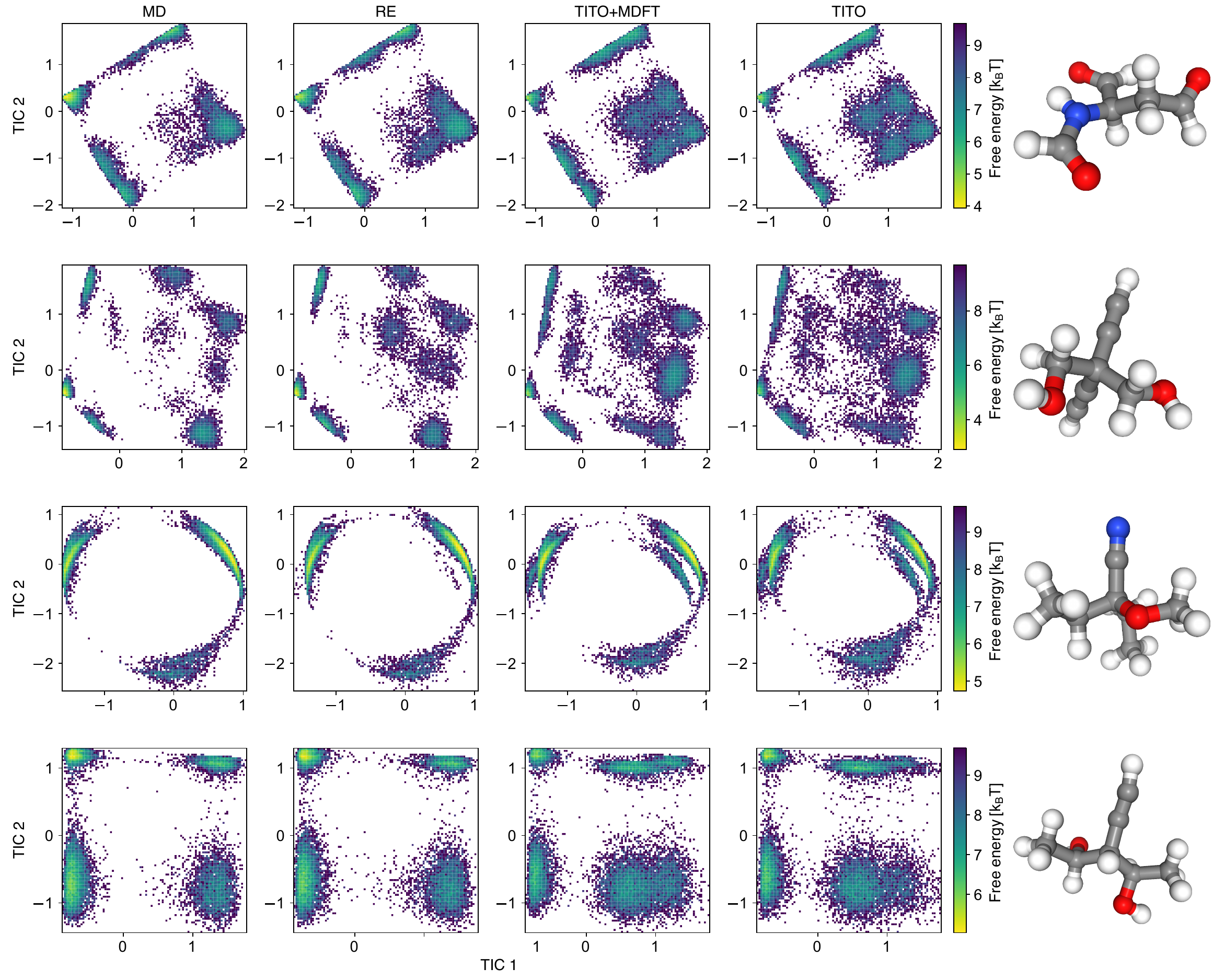}}
    \caption{Free energy landscape comparison between MD, RE and TITO+MDFT and TITO of test set molecules for which TITO predicts nanosecond meta-stable states missing or poorly sampled in RE.}
    \label{fig:si-meta-stable}
\end{figure}

\namedsubsection{Examples of nanosecond meta-stable state predictions of TITO missing in RE}
In Suppl.~Fig. \ref{fig:si-meta-stable} we collect 4 example test set molecules for which TITO samples states are not present in RE or MD, but are stable after 1 $\mathrm{ns}$ (per sample) ensemble simulation.

\begin{figure}[h!]
    \centering
    \includegraphics[width=\linewidth]{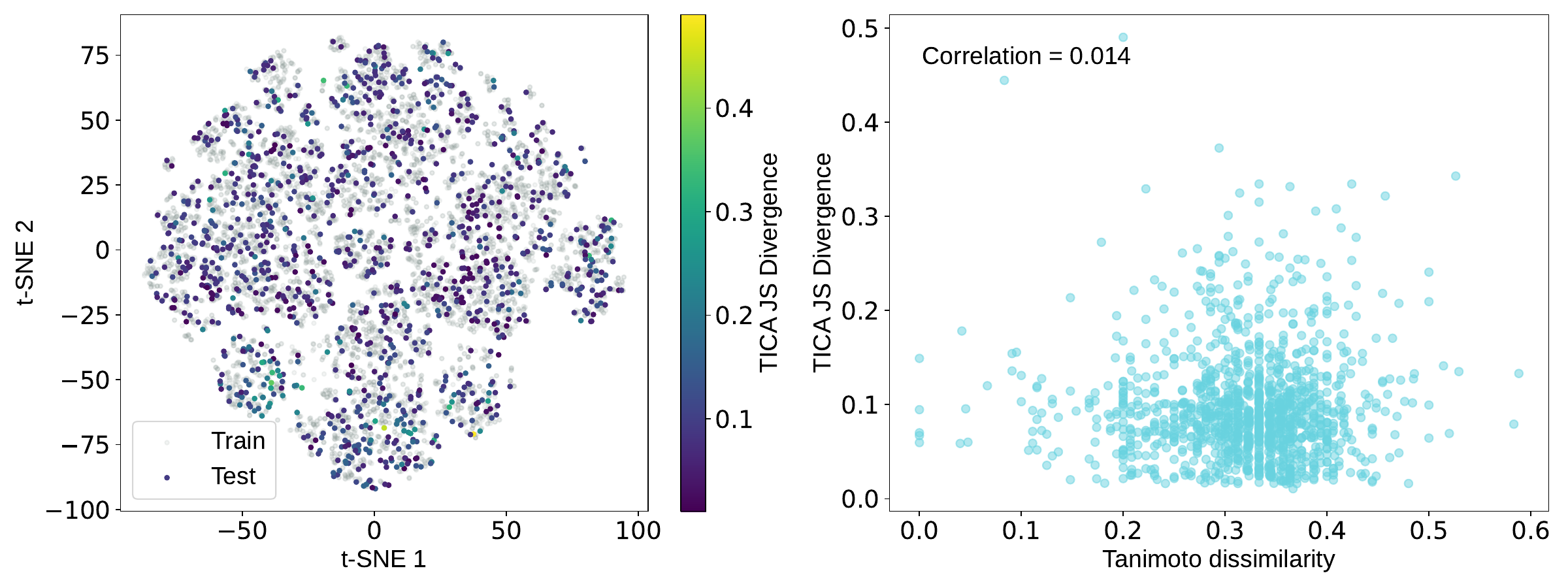}
    \caption{TICA Jensen-Shannon divergence vs chemical similarity for small molecules dataset. (Left) First two t-SNE projections for molecules in training and test set using Tanimoto dissimilarity as distance. Test set molecules are colored with Jensen-Shannon divergence of samples generated with TITO vs reference MD simulations. Test set is well covered by training set. (Right) Minimum Tanimoto dissimilarity of test set molecules w.r.t. training set vs Jensen-Shannon divergence. No correlation is observed and Tanimoto dissimilarities are low.}
    \label{fig:si-similarity-mdqm9}
\end{figure}

\begin{figure}

  \includegraphics[width=0.5\linewidth]{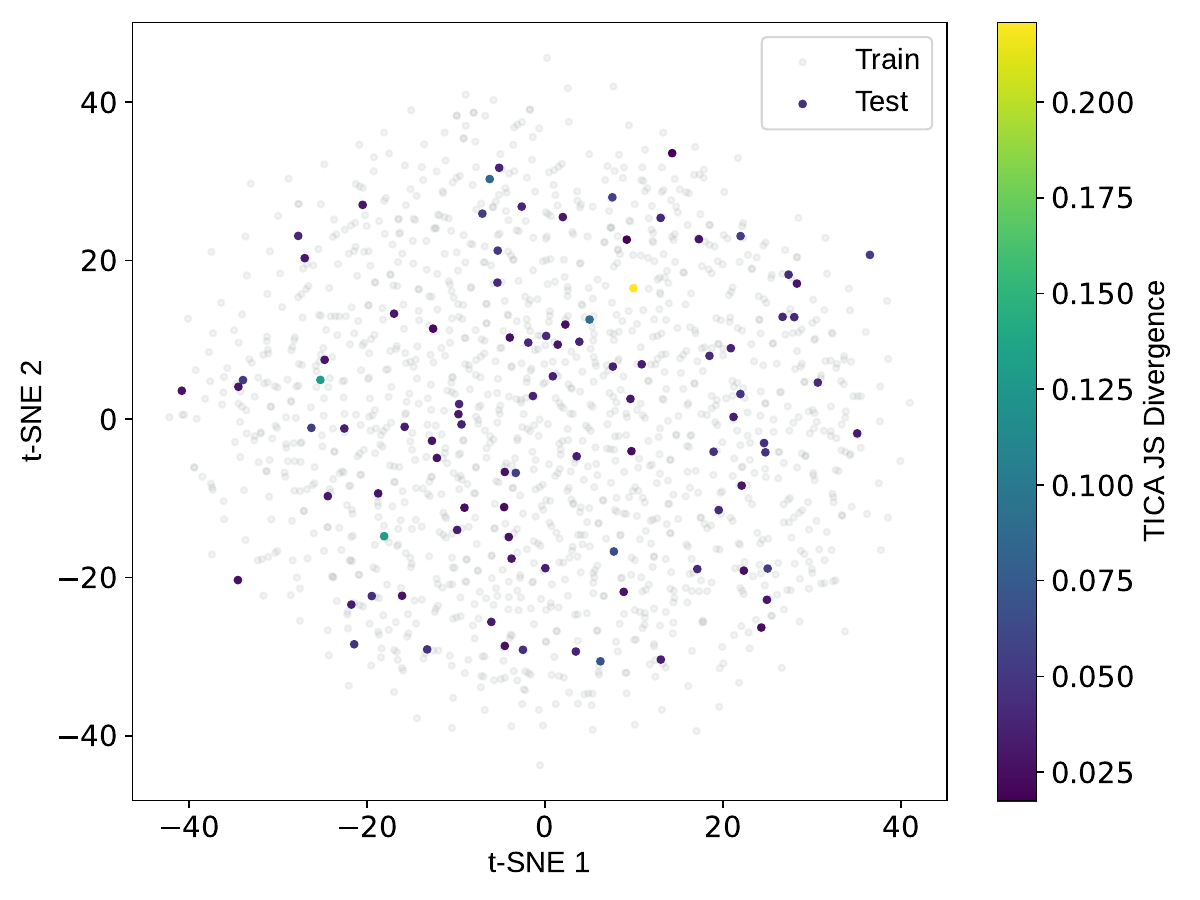}%
  \caption{TICA Jensen-Shannon divergence vs chemical similarity for tetra-peptides dataset. First two t-SNE projections for molecules in training and test set using Hamming distance. Test set molecules are colored with Jensen-Shannon divergence of samples generated with TITO vs reference MD simulations. Test set is well covered by training set.}%
  \label{fig:si-similarity-tw}
\end{figure}

\namedsubsection{Performance vs chemical similarity}
\label{sec:si-chamical-sim}
In Suppl.~Figs. \ref{fig:si-similarity-mdqm9} and \ref{fig:si-similarity-tw}, and Table \ref{table:si-similarity-tw} we show that equilibrium distribution errors do not correlate with chemical dissimilarity.

\begin{table}[h!]
   \begin{tabular}{|l|l|l|}
\hline
\textbf{Hamming distance}    & 0.25              & 0.75              \\ \hline
\textbf{Mean JS divergence}  & 0.043  & 0.040  \\ \hline
\textbf{\% of test peptides} & 64                & 26                \\ \hline
\end{tabular}

  \caption{Minimum Hamming dissimilarity of test set molecules w.r.t. training set vs Jensen-Shannon divergence. No correlation is observed}%
  \label{table:si-similarity-tw}
\end{table}

\begin{figure}[h!]
    \centering
    \includegraphics[width=\linewidth]{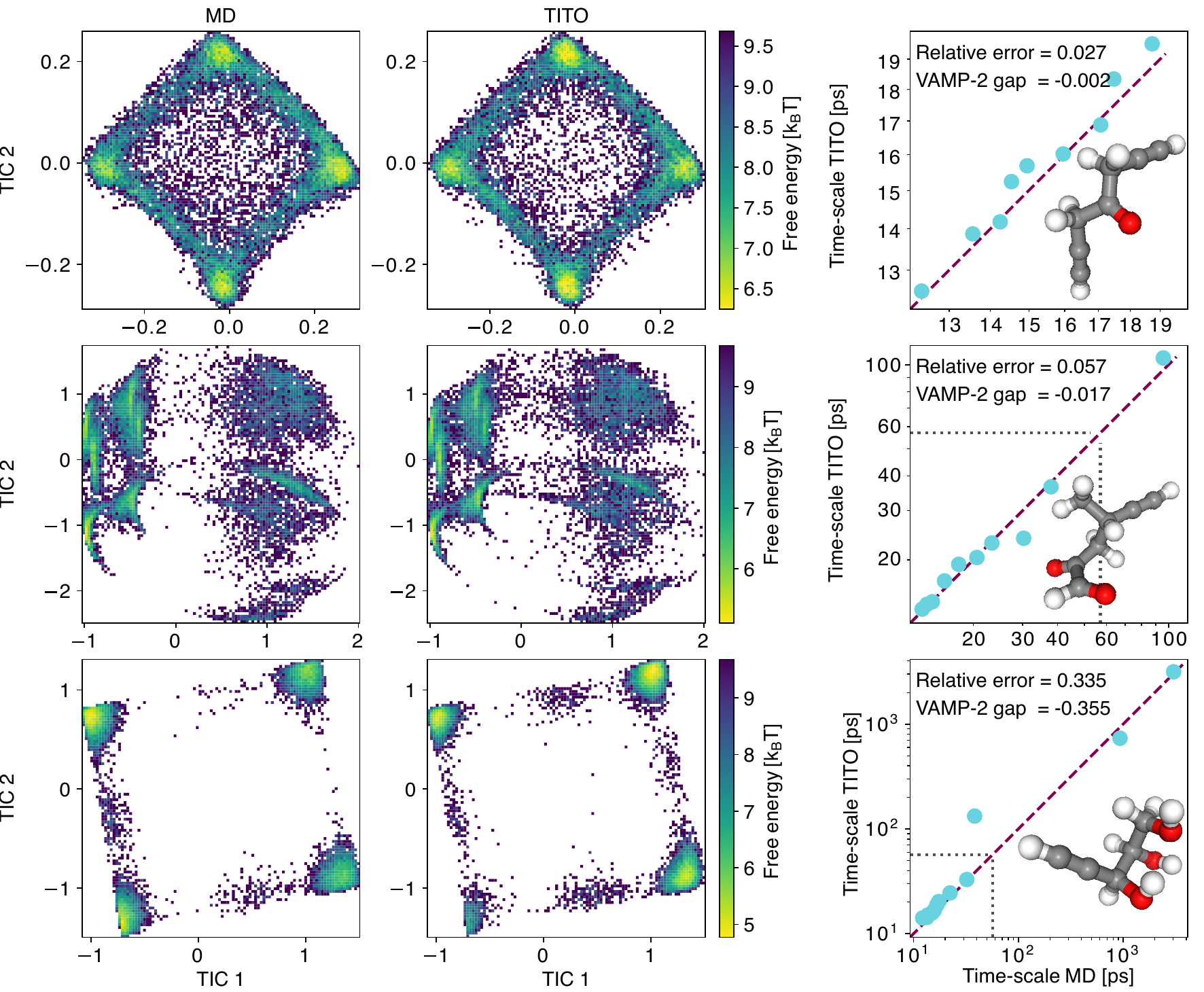}
    \caption{TITO accurately samples thermodynamic and kinetic  properties of molecules whose slowest process ranges from $\mathrm{ps}$ to $\mathrm{ns}$.}
    \label{fig:si-slow-and-fast}
\end{figure}

\namedsubsection{Accurately sampling different fast and slow molecules}
In Suppl.~Fig. \ref{fig:si-slow-and-fast} we show that TITO accurately samples thermodynamic and kinetic  properties of molecules whose slowest process ranges from $\mathrm{ps}$ to $\mathrm{ns}$.

\namedsubsection{TITO generates trajectories with multi-lag consistent timescales}

\begin{figure}[h!]
    \centering
    \includegraphics[width=0.5\linewidth]{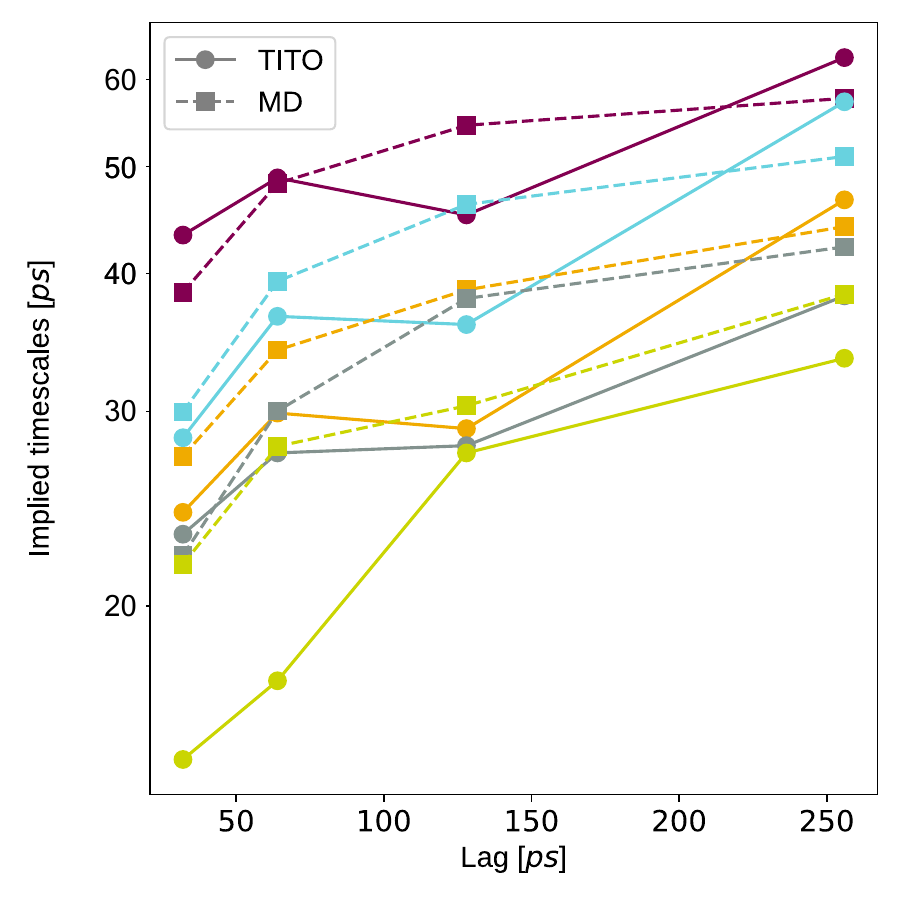}
    \caption{TITO samples faithful timescales w.r.t. MD, being consistent with the variational principle for Markov processes. VAMP implied timescales estimated at different lag times from MD and TITO. We show the 5 first VAMP timescales for the test set tetra-peptide studied in Fig. \ref{fig:marginals}.}
    \label{fig:si-ts-consistency}
\end{figure}

In Suppl.~Fig. \ref{fig:si-ts-consistency} we show that TITO generates trajectories with faithful VAMP timescales w.r.t. MD, exhibiting behavior consistent with the variational principle for Markov processes \cite{Nske2014}.

\begin{figure}[h!]
    \centering
    \includegraphics[width=0.75\linewidth]{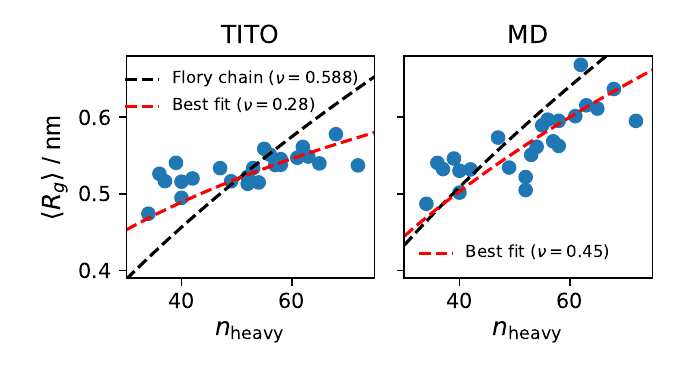}
    \caption{Scaling of radius of gyration $\langle R_g \rangle$ with number of heavy atoms in the pentapeptides, hexapeptides, heptapeptides and octapeptides. TITO results (left) are size extrapolations. MD values are computed on $100\,\mathrm n\mathrm{s}$ simulation trajectories for each of the extrapolation test systems. Dashed lines show ideal Flory chain (black) and best fit (red) scaling exponent $\nu$. }
    \label{fig:si-rg-extra}
\end{figure}

\begin{figure}[h!]
    \centering
    \includegraphics[width=1.0\linewidth]{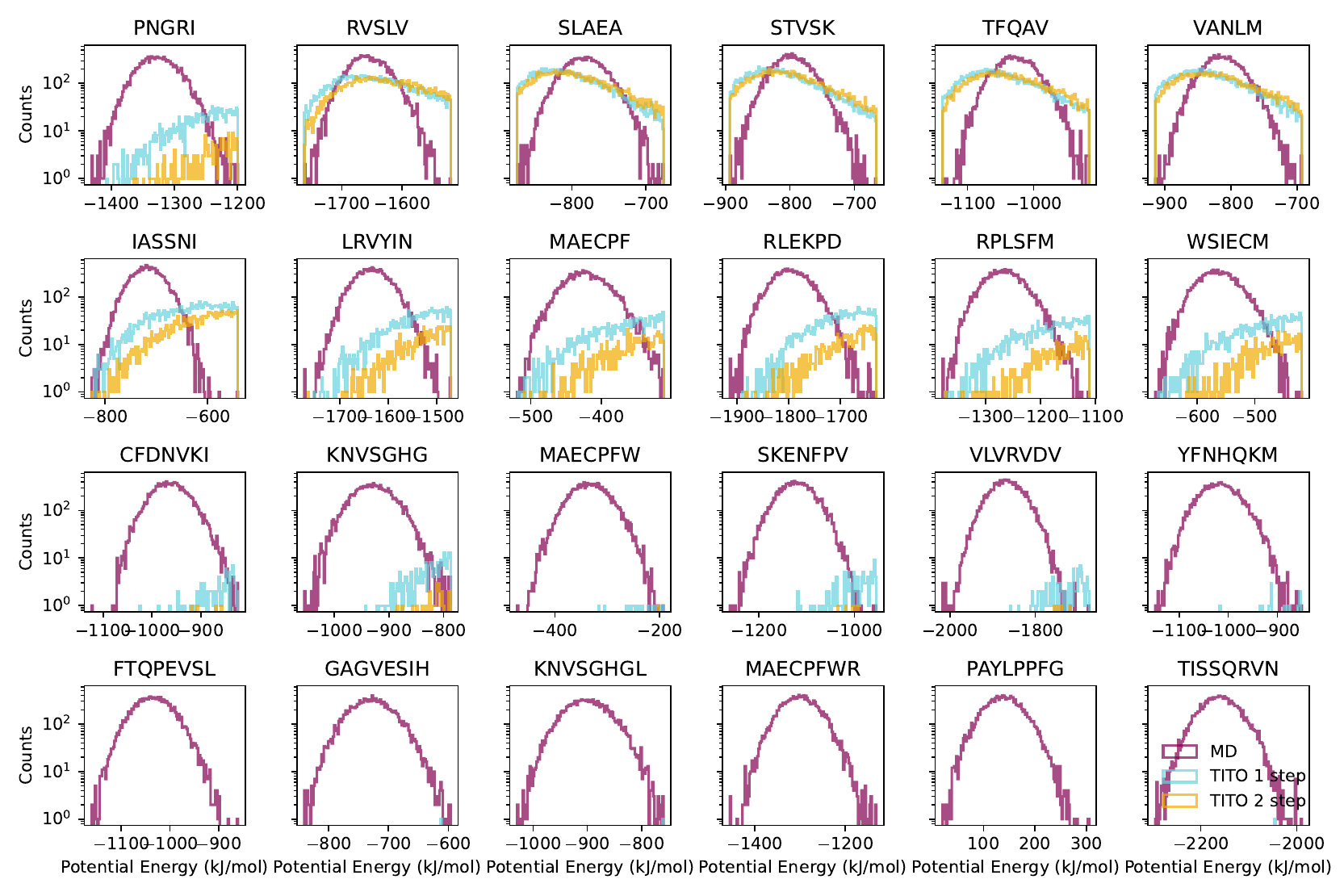}
    \caption{Potential energy distributions comparisons between MD (purple), single (blue) and two step (gold) sampling with TITO ($\Delta t=0.5\,\mathrm{ns}$) for pentapeptides, hexapeptides, heptapeptides and octapetides. Only samples in the range of sampled potential energies in the MD are included in the plot, if no samples are visible it means no samples generated fall within this potential energy range. }
    \label{fig:si-energy-extra}
\end{figure}

\namedsubsection{Peptide size extrapolation results}
Scaling of radius of gyration with number of heavy atoms (\ref{fig:si-rg-extra}) and potential energy distributions comparisons (\ref{fig:si-energy-extra}) for MD and TITO models trained on tetra-peptides in the pentapeptides, hexapeptides, heptapeptides and octapeptides.

\namedsubsection{Compute calibration example}

\begin{figure}[h!]
    \centering
    \includegraphics[width=1.0\linewidth]{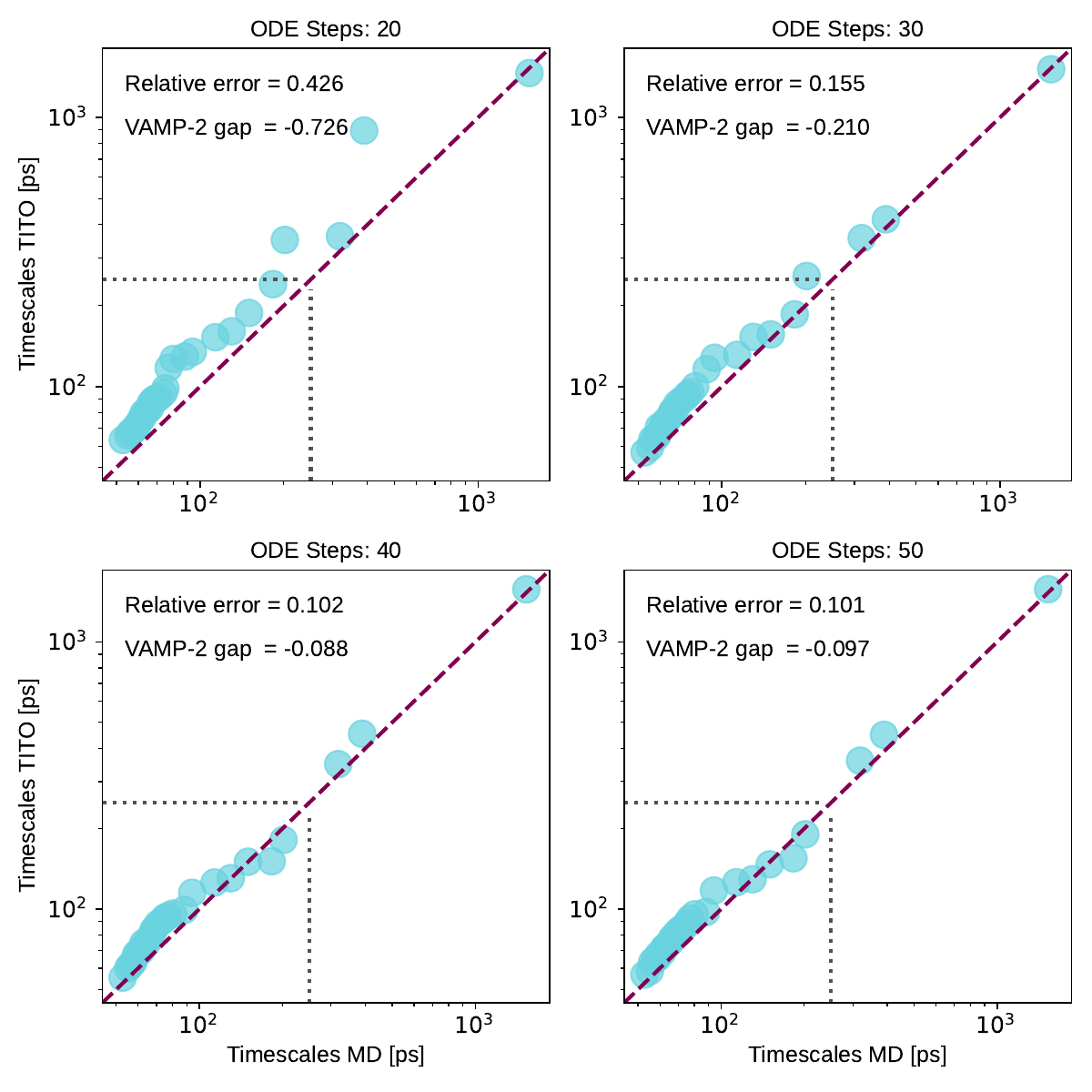}
    \caption{Correlation plot of MD and TITO timescales for different number of ODE steps. Error decrases with number of ODE steps.}
    \label{fig:si-compute-calibration}
\end{figure}

In Suppl.~Fig. \ref{fig:si-compute-calibration} we show how VAMP implied timescales agreement improves when increasing the number of ODE steps.

\namedsubsection{Experimental parameters}
\label{sec:si-experimental-parameters}
We show training and sampling parameters in Suppl.~Tables \ref{table:si-train-params}) and \ref{table:si-sampling-params}, respectively. When two numbers are shown separated by a forward slash (/), the first number refers to the small molecules dataset and the second to the tetra-peptides. 

\begin{table}[]
\begin{tabular}{|l|l|}
\hline
Layers              & 5          \\ \hline
Feature vector size & 64         \\ \hline
Embedding layers    & 2          \\ \hline
Learning rate       & 0.001\\ \hline
Batch size          & 750/32\\ \hline
Max lag             & 1 ns/ 5 ns \\ \hline
Lag distribution    & Uniform    \\ \hline
Training time    & 3 days    \\ \hline
Number of GPUs    & 4    \\ \hline
GPU model   & A100  \\ \hline
\end{tabular}
\caption{Training hyper-parameters of TITO models for small molecules/tetra-peptides.}
\label{table:si-train-params}
\end{table}

\begin{table}[]
  \begin{adjustbox}{width=\textwidth}
\begin{tabular}{c|c|c|c|c|c|}
\cline{2-6}
                                                     & Figure 2     & Figure 3 A and B & Figure 3 C        & Figure 4        & Figure 5 \\ \hline
\multicolumn{1}{|c|}{Lag}                            & 57 ps/250 ps & 1 ns             & 1 ns              & See figure      & 5 ns     \\ \hline
\multicolumn{1}{|c|}{Nested samples}                 & 640/500      & 1000             & 50,000            & 1 or 5 (nested) & 1        \\ \hline
\multicolumn{1}{|c|}{ODE steps}                      & 20/40        & 20               & 20                & 40              & 100      \\ \hline
\multicolumn{1}{|c|}{Integrator}                     & Euler        & Euler            & Euler             & Euler           & Euler    \\ \hline
\multicolumn{1}{|c|}{Batch size}                     & 32           & 128              & 32  & 50 000          & 51200    \\ \hline
\multicolumn{1}{|c|}{MD fine-tuning simulation time} & -            & -                & 10 ps             & -               & -        \\ \hline
\multicolumn{1}{|c|}{MD fine-tuning replicates}      & -            & -                & 32,000            & -               & -        \\ \hline
\multicolumn{1}{|c|}{Ultra long MD simulation time}  & -            & -                & 500 ns            & -               & -        \\ \hline
\multicolumn{1}{|c|}{Ultra long MD replicates}       & -            & -                & 32,000            & -               & -        \\ \hline
\end{tabular}
\end{adjustbox}
\caption{TITO sampling parameters for results in different figures in the main text for small molecules/tetra-peptides. }
\label{table:si-sampling-params}
\end{table}

\namedsubsection{Maximum throughput details}
\label{sec:si-max-thoughput}
For both datasets, we computed selected a random molecules with the average number of heavy atoms in the dataset. We increased the batch size until we saturated the GPU RAM, and sampled with the maximum possible time lag for 24hrs.







\end{document}